\documentclass[twocolumn,english,aps,prl,longbibliography,floatfix,superscriptaddress]{revtex4-1}

\usepackage{natbib,bm,graphicx,url,epsfig,psfrag}
\usepackage{graphicx} 
\usepackage{dcolumn}
\usepackage{bm}
\usepackage{amsmath}
\usepackage{amssymb}
\usepackage{xcolor}
\usepackage[caption=false]{subfig}
\usepackage{bbold}
\usepackage{bm}
\usepackage{mwe}
\usepackage[normalem]{ulem} 
\usepackage{tikz}
\usepackage{xfrac}
\usepackage{blindtext}
\usepackage{times}
\usepackage{placeins}
\newcommand{\redout}[1]{{}}
\renewcommand{\vec}[1]{{\boldsymbol #1}}

\usepackage[colorlinks=true,linkcolor=blue,urlcolor=blue,citecolor=blue]{hyperref} 

\definecolor{darkblue}{HTML}{004D6B}
\definecolor{darkred}{HTML}{8c1515}

\usepackage{hyperref}
\hypersetup{
	colorlinks=true,
	urlcolor=darkred,
	citecolor=darkblue,
	linkcolor=darkred,
	breaklinks
}

\begin{document}

\title{Thermal Hall response: violation of gravitational analogues and Einstein relations}
\begin{abstract}
The response of solids to temperature gradients is often described in terms of a gravitational analogue:
the effect of a space-dependent temperature is modeled using a space dependent metric.
We investigate the validity of this approach in describing the bulk response of quantum Hall states and other gapped chiral topological states. 
To this end, we consider the prototypical Haldane model in two different cases of (i) a space-dependent electrostatic potential and gravitational potential and (ii) a space-dependent temperature and chemical potential imprinted by a weak coupling to non-interacting electron baths or phonons.
We find that the thermal analogue is \textit{invalid};
while a space dependent gravitational potential induces transverse energy currents proportional to the third derivative of the gravitational potential, the response to an analogous temperature profile vanishes in limit of weak coupling to the thermal bath. Similarly, the Einstein relation, the analogy between the electrostatic potential and the internal chemical potential, is not valid in such a setup.
\end{abstract}
\author{Jinhong Park}
\affiliation{Institute for Theoretical Physics, University of Cologne, Z{\"u}lpicher Str. 77, 50937 K{\"o}ln, Germany}
\author{Omri Golan}
\affiliation{Department of Condensed Matter Physics, Weizmann Institute of Science, Rehovot 76100, Israel}
\author{Yuval Vinkler-Aviv} 
\affiliation{Institute for Theoretical Physics, University of Cologne, Z{\"u}lpicher Str. 77, 50937 K{\"o}ln, Germany}
\author{Achim Rosch}
\affiliation{Institute for Theoretical Physics, University of Cologne, Z{\"u}lpicher Str. 77, 50937 K{\"o}ln, Germany}
\date{\today}
\maketitle

{\it Introduction.---}
Thermal transport in topological matter has been of high interest as its quantization can reveal the topological nature of the underlying state of matter~\cite{Kane1997,Cappelli2002}. Recently, a half-integer quantized thermal conductance has been measured in $\nu = 5/2$ fractional quantum Hall state~\cite{Banerjee2018} and $\alpha$-$\rm {RuCl_3}$~\cite{Kasahara2018,bruin2021,Yokoi2021}; the latter is a candidate for realistic materials of chiral Kitaev spin liquid~\cite{Kitaev2006}. Here a half-integer thermal conductance can be viewed as a smoking-gun signature of the existence of gapless chiral Majorana fermion on the edges. While such thermal edge transport in topological materials is fairly well understood (e.g., for a chiral spin liquid~\cite{Kitaev2006,Vinkler2018,Ye2018}), topological response in the bulk and the corresponding bulk-boundary correspondence is actively discussed~\cite{Bradlyn2015, Nakai2017,Kapustin2020,Huang2021}. 

In contrast with charge transport, where 
the electrostatic potential is coupled to the electron density and thus the transport coefficients can be derived from the linear response theory, there is no apparent term added in the Hamiltonian for the thermal transport. Luttinger~\cite{Luttinger1964} suggested that thermal transport can be investigated via the coupling to fictitious and spatially varying metric tensor (i.e., gravity). In this case the Hamiltonian is given by
\begin{align}
H[\psi]=\int\! d^d\vec r\, \sqrt{g(\vec r)} \,h(\vec r)=\int \! d^d\vec r\, (1+\psi(\vec r))\, h(\vec r)
\end{align}
where $h(\vec r)$ is the (flat-space) energy density and the gravitational potential $\psi(\vec{r})$ describes how the metric varies spatially.

Roughly, the analogy of transport in curved space with the problem of a space-dependent temperature $T(\vec r)$ is obtained when one considers
the density matrix
\begin{align}
\rho \sim  e^{- \int \! d^d\vec r\, \beta(\vec r) h(\vec r)}\label{rhoW}
\end{align}
with a space-dependent inverse temperature $\beta(\vec r)=1/T(\vec r)$. Comparing this to $e^{-\overline{\beta} H[\Psi]}$ suggests to
identify
\begin{align}
\frac{1}{T(\vec r)} = \beta(\vec r)= \overline{\beta} (1+\psi(\vec r)).\label{eq:analogy}
\end{align}
where $\overline{T}=1/\overline{\beta}$ is a reference temperature. While this analogy is appealing, it is also rather obvious that it can only be of limited validity. Importantly, any problem with a space-dependent temperature is by definition a non-equilibrium problem. Thus the density matrix~of~Eq.~\eqref{rhoW} does {\em not} describe the steady state of the system. 

\begin{figure}
    \centering  \includegraphics[width=0.45\columnwidth]{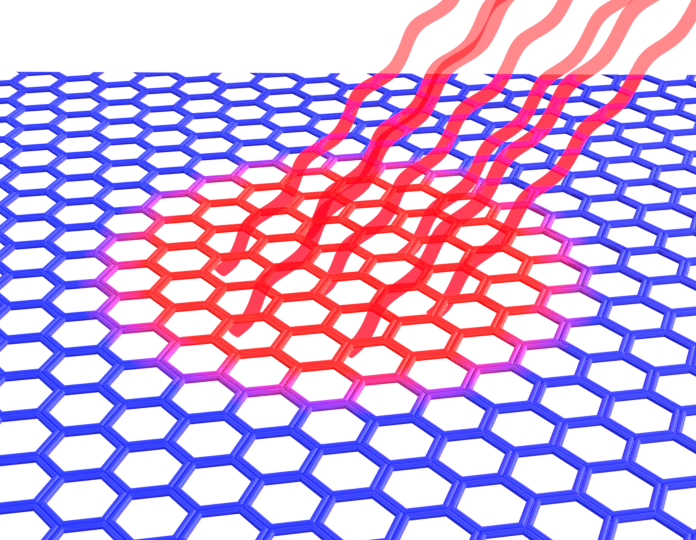}\hspace{0.07\columnwidth}
                \includegraphics[width=0.45\columnwidth]{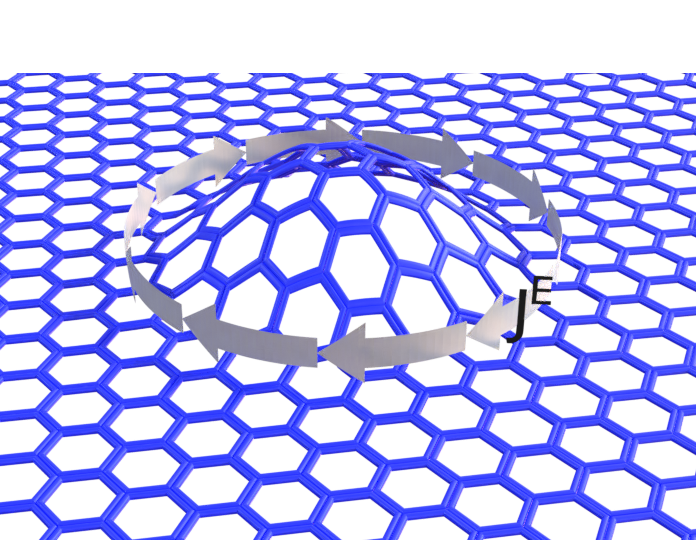}
    \caption{
     {\bf Two systems for the thermal Hall bulk response.}
     (a) A focused laser beam locally heats the system via phonons and creates a temperature profile $T(r)$. (b) Using a gravitational analogue, this would correspond to a distorted lattice with space dependent hoppings. In a Chern insulator, the curvature induces a circulating energy current, while 
    the temperature bump has no such effect.}
    \label{Fig1:tempgravbump}
\end{figure}
The gravitational analogy is well established for the calculation of thermal transport in the thermodynamics limit and has, for example, been used to classify topological matter \cite{Ryu2012}. 
According to the so-called {\it Luttinger relation}, 
 the thermal conductivity tensor $\kappa$ describes both the response to gradients of $\psi$ and $T$
\begin{align}
\vec{J}^E=-\kappa (T \vec \nabla \psi + \vec \nabla T )\label{eq:luttingerR}
\end{align}
where $\vec{J}^E$ is the energy current density. Within linear response theory, $\kappa$ is therefore routinely calculated by considering the response to a space- and time-dependent $\psi\sim e^{i (\vec q\vec r-\Omega t)}$. Importantly, one has to use the `transport limit' for such a calculation by taking first the limit $\vec q \to 0$ and only then $\Omega \to 0$. In case that both $\psi$ and $T$ are present, one has to identify $T(\vec r)$ with the internal temperature calculated from the local energy density rather than the thermodynamic temperature\redout{, see Ref.~\cite{Cooper1997}}~\cite{Cooper1997}. 
The Luttinger relation is in close analogy to the Einstein relation for electric transport, 
\begin{align}
\vec{J}^C =\sigma \vec \nabla  (\phi + \mu),\label{eq:einsteinR}
\end{align}
 where $\phi$ is an external potential and $\mu$ the internal chemical potential calculated from the local density (and not the electrochemical potential).
In the case of broken time-reversal symmetry, one has to be careful when defining the correct `transport currents'
which have to be distinguished from equilibrium currents related to the magnetization of the sample, see Refs.~\cite{Cooper1997, Qin2011, Bradlyn2015,Gromov2015, Kapustin2020,Huang2021}.

In a quantum Hall system both $\sigma_{xy}$ and $\kappa_{xy}/T$ are quantized while the longitudinal conductivites vanish, $\kappa_{xx}=\sigma_{xx}=0$.
There is, however, a remarkable difference in the response to a static, space-dependent
 electrostatic potential $\phi(\vec r)$ and a gravitational potential $\psi(\vec r)$. For example, $\phi(\vec r)$ may arise from an electric charge close to the surface of a topological insulator~\cite{Qi2009}. In this case, the potential induces circulating currents perpendicular to the potential gradients which can be computed directly from Eq.~\eqref{eq:einsteinR}. Remarkably, this is {\em not} the case when the response to $\psi(\vec r)$ is calculated which may arise\redout{, e.g.,} due to a bump in the 2d material, see Fig.~\ref{Fig1:tempgravbump}(b). In the context of relativistic field theories, the bulk response is 
given by~\cite{Stone2012}
\begin{align} \label{eq:relativisticresponse}
    J^E_i (\vec{r}) &= \frac{\hbar c^2 (c_R-c_L)}{96 \pi} \epsilon^{ij} \partial_j R \nonumber \\
    &\approx-\frac{ \hbar c^2 (c_R-c_L)}{48 \pi} \epsilon^{ij} \partial_j \nabla^2 \psi
\end{align} 
where $c$ is the speed of light, $R =-2\nabla^2 \psi + O (\psi^2)$ the curvature, and $c_R-c_L$ the difference of left-moving and right-moving central charge characterizing the edge modes of the system. $c_R-c_L$ is directly related to the quantized thermal Hall conductivity
\begin{align}
\frac{\kappa_{xy}}{T}=(c_R-c_L)\frac{\pi^2 k_B^2}{3h}.
\end{align} 

Remarkably, Eq.~\eqref{eq:relativisticresponse} predicts that the thermal topological response to a gravitational potential is proportional to 
the third derivative of $\psi(\vec r)$ while the Luttinger relation suggests a response Eq.~\eqref{eq:luttingerR} proportional to the first derivative.
This is not a direct contradiction because Eq.~\eqref{eq:relativisticresponse} has been calculated for a smooth {\em static} potential, i.e., by taking first the limit $\Omega \to 0$, while Eq.~\eqref{eq:luttingerR} is valid in the opposite limit where one first considers the limit $\vec q \to 0$. The topological bulk response Eq.~\eqref{eq:relativisticresponse} is directly linked to the gravitational anomaly~\cite{AlvarezGaume1984} of the edge 
theory: an apparent violation of energy conservation at the edge in the presence of gravitational potentials can be explained by the inflow of energy from the bulk~\cite{Stone2012, Golan2018}.

An interesting observation from Eq.~\eqref{eq:relativisticresponse} is that the topological bulk response of relativistic theories is proportional to $c^2$ which immediately suggests that the effect cannot be fully universal in non-relativistic topological phases where it is unclear what should replace the speed of light.
In this context, it would be desirable to understand the bulk response for non-relativistic theories realized in all condensed matter settings.

In this paper, we investigate whether the gravitational analogy and Luttinger relations can be used to calculate the response to space-dependent temperature profiles which arise when
a system is heated locally (see Fig.~\ref{Fig1:tempgravbump}(a)). More specifically, we will show that the gravitational analogy does {\em not} hold for temperature profiles $T(\vec r)$
even in a regime where $T(\vec r)$ is much smaller than the gap of the system. Similar statements hold for the electric case. Furthermore, we argue that the bulk response to local gravitational potentials is not quantized in non-relativistic theories.

\begin{figure} 
\centering
\includegraphics[width=.95\columnwidth]{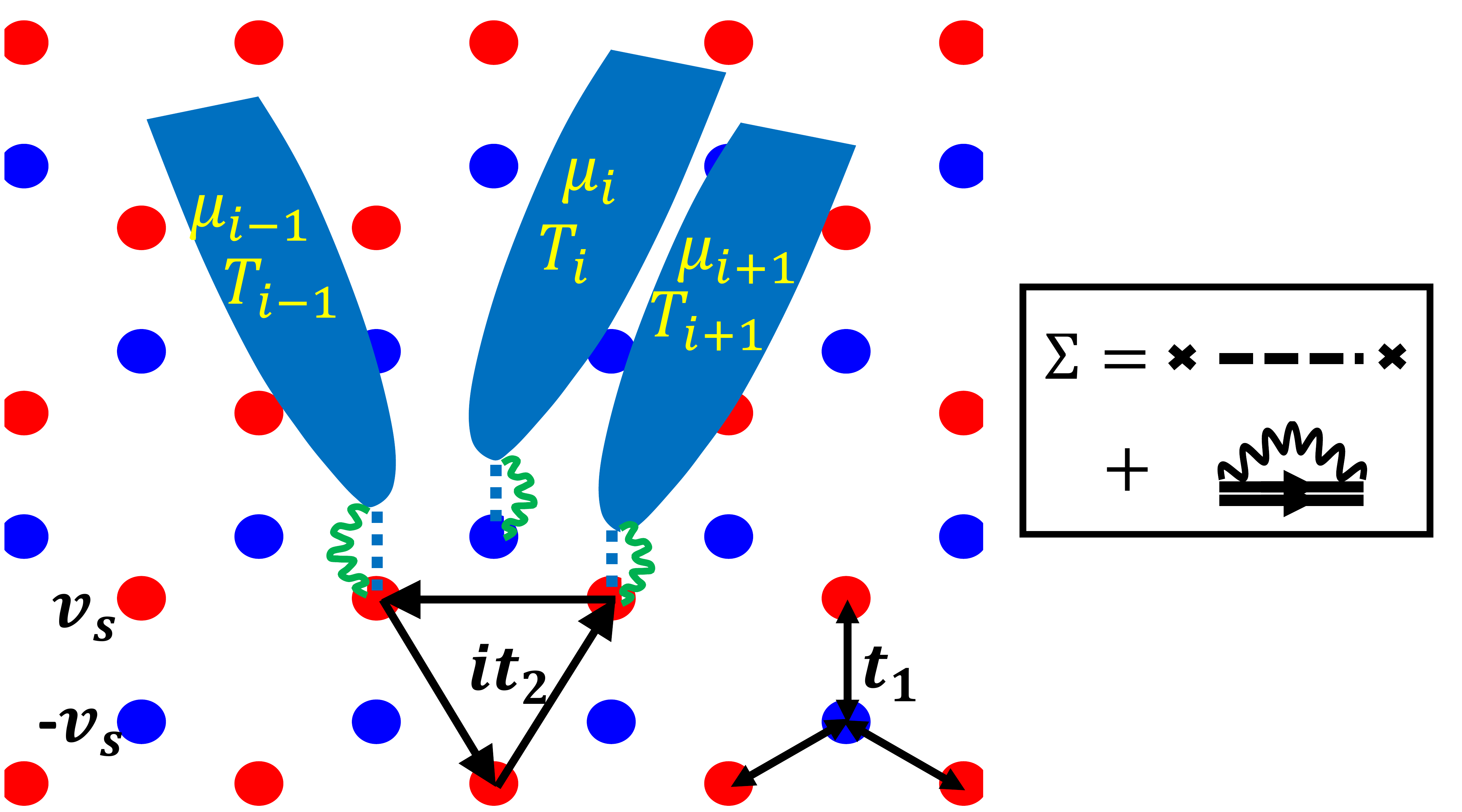}
\caption{{\bf Schematic drawing of the model.} Each lattice point $i$ in the Haldane lattice model (cf. Eq.~\eqref{eq:haldane}) is weakly coupled to both a phonon bath (wiggly line) with temperature $T_i$ and to an electronic wire with chemical potential $\mu_i$ and the same  temperature $T_i$ via a tunnel contact (dashed line). On the right hand side we show the corresponding Keldysh self-energy diagrams which are evaluated self-consistently.
}
\label{Fig2:systemattachedlead}
\end{figure}

{\em Gravitational response.---} As a concrete example, we consider the Haldane model~\cite{Haldane1988}, which describes a Chern insulator, defined on a honeycomb lattice.
\begin{align}
\hat{H}[\psi]=-\sum_{i, j} t_{ij} (1+\psi(\vec r_{ij})) c^\dagger_i c_j+ \sum_{i} v_i (1+\psi(\vec r_{i})) c^\dagger_i c_i \label{eq:haldane}
\end{align}
where $t_{ij}$ encodes a real-valued nearest-neighbor hopping $t_1$ and a purely imaginary next-nearest neighbor hopping
$\pm i t_2$ (see Fig.~\ref{Fig2:systemattachedlead}). $v_i=\pm v_s$  is a staggered potential.
To model the effect of a gravitational potential, all terms in the Hamiltonian depend on a smoothly varying gravitational potential 
$\psi(\vec{r}_{ij})$ with $\vec r_{ij}=(\vec r_i + \vec r_j)/2$.

At  $v_s < 3\sqrt{3} t_2$, the system is in a topological phase with Chern number $c_{R} - c_{L} = -1$.  At the quantum phase transition to the trivial phase at $v_s =3\sqrt{3} t_2$, the gap closes at the $K$ while it remains finite at the $K'$ point~\cite{Haldane1988}. 
Close to this transition, the system is accurately described by its continuum limit
\begin{align} \label{Haldanemodel}
\hat{H}[\psi]&\approx \hat{H}_c[\psi]=\int (1+\psi(\vec r))  \hat{h}_c(\vec r) d^2 \vec r \\
\hat{h}_c(\vec r) &=  \Psi^{\dagger} (\vec{r}) \left (- i v \vec{\sigma} \cdot \nabla_{\vec{r}} + (M - \lambda^2 \nabla_{\vec{r}}^2 )\sigma_z \right) \Psi (\vec{r}). \nonumber
\end{align}
with $M=v_s-3\sqrt{3} t_2$, $v = -\sqrt{3} t_1 a/2$, $\lambda^2=3 \sqrt{3} t_2 a^2/4$. Here $a$ is the lattice constant. The two-component spinor $\Psi^\dagger=(\Psi^\dagger_1,\Psi^\dagger_2)$ creates electrons close to the $K$ point.

Employing this continuum model, we first consider the response to a gravitational potential $\psi(\vec r)$ and an electrostatic potential $\phi(\vec r)$ at zero temperature and for vanishing chemical potential.
For simplicity, it is assumed that $\psi (\vec{r})$ and $\phi (\vec{r})$ vary only in the $x$ direction, but are constant in the $y$ direction. 

For the calculation of the gravitational response, we first define the energy current density operator $\hat{\vec{J}}{}^{E}_{\psi} (\vec{r})$ in the presence of $\psi (\vec{r})$. 
$\hat{\vec{J}}{}^E_{\psi} (\vec{r})$ can be uniquely determined by (i) requiring the continuity equation $\nabla_{\vec{r}}\cdot\hat{\vec{J}}{}_{\psi}^{E} (\vec{r}) = i  [ (1 + \psi (\vec{r})) \hat{h}_c (\vec{r}), \hat{H}_c [\psi] ] / \hbar$ 
and (ii) imposing that $\hat{\vec{J}}{}_{\psi}^{E}$ is related to the zero-potential energy current operator $\hat{\vec{J}}{}^{E}$ as $\hat{\vec{J}}{}_{\psi}^{E} = (1 + \psi)^2 \hat{\vec{J}}{}^{E}$~\cite{Cooper1997,Qin2011,Vinkler-Aviv2019}. The calculation is done most conveniently in momentum space, where 
$\hat{\vec{J}}{}^{E} (\vec{q})$ has a form  $\hat{\vec{J}}{}^{E} (\vec{q}) = \int d\vec{K}/(2\pi)^2 \Psi^{\dagger}_{\vec{K}-\vec{q}/2}
 \vec{J}{}^E_{\vec{K}- \vec{q}/2; \vec{K} +\vec{q}/2} \Psi_{\vec{K}+\vec{q}/2}$. 
Using standard linear-response theory,  the expectation value of the energy current operator $\vec{J}_{\psi}^{E} (\vec r) \equiv  \langle  \hat{\vec{J}}{}_{\psi}^{E} (\vec r) \rangle$ can be obtained in the linear order of the {\em static} gravitational potential $\psi$ (see the supplemental material~\cite{Supple} for details).
Expanding in $\vec q$ for smoothly varying $\psi(x)$ we obtain
\begin{align} \label{eq:Haldanenonrelativisticresponse}
J_{y, \psi}^{E} (\vec r) & \approx \frac{  \partial_x^3\psi (\vec{r}) }{96 \pi} 
\bigg [ 4\lambda^2 M \theta (-M) \nonumber 
\\
& +  v^2 \left \{\textrm{sgn} (M)  -3 
\log \left (\frac{4 e^{-5/3} K_{\rm{cut}}^2\lambda^4}{v^2 + 4 \lambda^2 M \theta (M)} \right ) \right \}\bigg].
\end{align}
where we used that $\psi(\vec r)=\psi(x)$ depends only on the $x$-coordinate in our setup.
Here $K_{\rm{cut}}$ is a ultra-violet momentum cutoff which is needed to obtain a finite result.

Two main conclusions can be drawn from Eq.~\eqref{eq:Haldanenonrelativisticresponse}.
(i) We have confirmed that the response to a {\em static} gravitational potential is {\em not} proportional  $\partial_x \psi$ as suggested by the Luttinger relation, Eq.~\eqref{eq:luttingerR},
(which has been derived for a time-dependent potential in the `transport limit'). 
It is instead proportional to $\partial_x^3 \psi$ as suggested by the relativistic anomaly formula, Eq.~\eqref{eq:relativisticresponse}.
(ii) The prefactor of the anomaly response is, however, not simply given by $\frac{\hbar c^2}{48 \pi}$ (cf. Eq.~\eqref{eq:relativisticresponse}). Instead it is non-universal and depends on the microscopic parameters and the cutoff in a non-universal way. As discussed above, this is not completely unexpected as the relativistic formula, Eq.~\eqref{eq:relativisticresponse}, depends on a dimension-full quantity, the speed of light.
We have checked that the same calculation which results in Eq.~\eqref{eq:Haldanenonrelativisticresponse}, leads to $J_{y,\psi}^{ E} (\vec r) = \frac{\hbar c^2}{96 \pi} \partial_x^3  \psi (\vec{r}) \textrm{sgn} (M)$ when one uses a fully relativistic model ($\lambda=0$, $K_{\rm{cut}}=\infty$),
thus $(c_R-c_L)=\frac{1}{2} \textrm{sgn} (M)$ consistent with the half-integer Chern number of the fully relativistic model. 

 In contrast, a calculation for the charge response to $\phi (\vec{r})$ results 
\begin{align} 
J_{y, \phi}^{C} (\vec{r}) = e \partial_x \phi (\vec{r}) \theta (-M)/(2 \pi \hbar)\label{eq:Jc}
\end{align}
 as expected (see the supplemental material~\cite{Supple}). In this case the response is linear in gradient consistent with Eq.~\eqref{eq:einsteinR}, the prefactor is fully universal and given by the topological response of the Haldane model characterized by $\sigma_{xy}=\frac{e^2}{2 \pi \hbar}$.


\begin{figure}[t]
\includegraphics[width=\columnwidth]{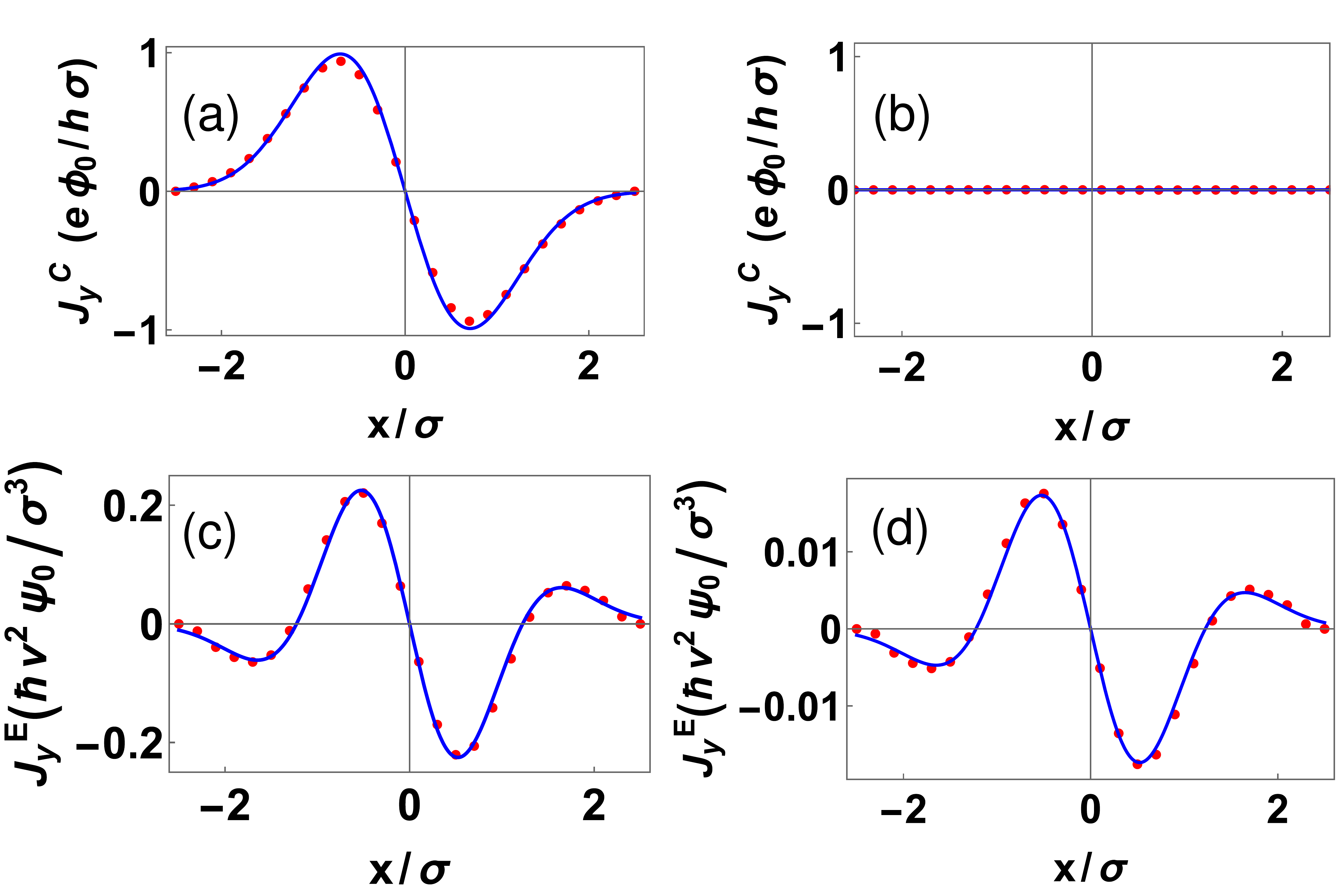}
\caption{{\bf Hamiltonian response.} (a-b) The electrical Hall current $J_{y, \phi}^{C}$  and (c-d) the energy Hall current $J_{y, \psi}^E$  in response to electrostatic potential $\phi (x) = \phi_0 \exp (-x^2/\sigma^2)$ and gravitational potential $\psi (x) = \psi_0 \exp (-x^2/\sigma^2)$, respectively. Left: topological phase, right: trivial phase.
Red dots: numerical calculation using the Haldane lattice model (see the supplemental material~\cite{Supple}). Blue lines: analytical result using Eq.~\eqref{eq:Jc} for the electrical response. For the gravitational response a  fit to $J_{y, \psi}^E = C \frac{v^2}{48 \pi} \partial_x^3 \psi (x)$ is shown  with $C \approx 8.7$  for (c) and $C \approx 0.67$ for (d). Parameters:  $\sigma = 10 a$, $\phi_0=\psi_0=0.1$, $t_1  = t_2=1$ and the staggered potential $v_s  =(3 \sqrt{3} - 1)t_2$ in the topological phase (left) and $v_s  =(3 \sqrt{3}+ 1)t_2$  in the trivial phase (right).
}
\label{fig:Mechresponse}
\end{figure}

In Fig.~\ref{fig:Mechresponse} we show the charge and heat currents~\cite{footnote} calculated directly from the lattice model~\eqref{eq:haldane} in response to an electric and a gravitational potential, respectively. The numerics confirm that the responses are proportional to $\partial_x \phi$ and $\partial^3_x \psi$~\cite{Note}.
In the electric case the response is only finite in the topological phase and the prefactor matches exactly the universal result of Eq.~\eqref{eq:Jc}. The gravitational response, $J_{y,\psi}^E = C \frac{v^2}{48 \pi} \partial_x^3 \psi (x)$, is 
non-universal, with $C \approx -8.7$ in the topological and $C \approx -0.67$ in the trivial phase for the chosen parameters. The result from the continuum model~\cite{Supple}, depends strongly on the cutoff with $C\approx -12$ and $C \approx -4.4$ for $K_{\rm cut}=2 \pi$ ($C\approx -9.8$ and $C \approx -2.3$ for $K_{\rm cut}=\pi$),
but is roughly consistent with the lattice calculation.

{\em Response to temperature bump.---}
We next consider the temperature bump $T (\vec{r})$ and the chemical potential bump $\mu (\vec{r})$ in the absence of $\psi (\vec{r})$ and $\phi (\vec{r})$. 
To be able to change locally the temperature and the chemical potential, we couple weakly to each lattice site $i$ of a Chern insulator \cite{Supple}, both a bath of phonons with temperature $T_i$ and a wire with a chemical potential $\mu_i$ and the same temperature $T_i$, see Fig.~\ref{Fig2:systemattachedlead}, using a tunneling contact of strength $V$.
The coupling is described by $\hat{H}_t+\hat{H}_{\rm ph}$ with
\begin{align}
\hat{H}_t&=\sum_{i,q}    \epsilon_q d^\dagger_{i, q} 
d_{i, q} + V d^\dagger_{i, q}c_{i}  + h.c.  \nonumber \\ \label{phononcoupling}
\hat{H}_{\rm{ph}}&=\sum_{i,  q} \omega_q a^\dagger_{i,  q} 
a_{i,  q} + g\, c_{i}^{\dagger} c_{i }
\left (a^{\dagger}_{i, q} + a_{i, q} \right),
\end{align}
 where we parametrize the (ohmic) phonon coupling by the parameter $\alpha$ with  $\pi g^2  \sum_q \delta(\omega-\omega_q)=\alpha \omega $
and the tunnel coupling by $\Gamma=\pi V^2 \sum_q  \delta(\omega-\epsilon_q)$. For simplicity we assume that both $\alpha$ and $\Gamma$ are $\omega$ indepdendent. The information on $T_i$ ($T_i$ and $\mu_i$) is encoded in the $i$-dependent Bose function (Fermi functions) used to describe the occupation of the phonons (fermions). Within our model, the phonon baths are strictly local and thus unable to transport heat, which simplifies the analysis of heat currents. 


%
As we are studying now a non-equilibrium state, we use the Keldysh formalism. The attached wires are treated exactly, while we use a self-consistent one-loop approximation for the phonons, see Fig.~\ref{Fig2:systemattachedlead}. This is equivalent to the solution of a corresponding quantum-Boltzmann equation~\cite{Rammer1986}. We use $\mu_i$ and $T_i$ which are translationally invariant in the $y$ direction. The system is infinite in the $y$ direction, while we use either 16 or 32 sites in the $x$-direction with periodic boundary conditions.
%
In the following, we assume (i) that all temperatures and chemical potential are always much smaller than the gap, $|\mu_i|, T_i \ll |M|$, and (ii) that they vary on a length scale $\sigma$ {\em larger} than both the correlation length $\xi \sim v/|M|$ of the gapped system and the phonon-induced mean-free path $\xi_{\rm ph}$ of thermal excitations, $\sigma \gg \xi, \xi_{\rm ph}$.

To determine the local temperature and the local chemical potential imprinted on our Chern insulator at position $\vec r$,
we have to calculate the local distribution function  defined by $f^{\text{eff}}_{\vec r}(\omega)=G^{<}_{\vec r, \vec r}(\omega)/(G^R(\omega)-G^A(\omega))$ with the local Green functions $G^{R/A}(\omega)$. As the Green functions decay on the length scale $\xi$  for $|\omega| \ll |M|$ and on the length scale $\xi_{\rm ph}$ for $|\omega| \gtrsim |M|$,  the distribution function is effectively averaged over these length scales.
As $\sigma \gg \xi, \xi_{\rm ph}$, the local temperature \cite{Lenarcic2018} and chemical potential of the Chern insulator are thus well defined and 
determined by the attached wires and phonons, $f^{\text{eff}}_{\vec r_i}(\omega)\approx 1/(e^{(\omega-\mu_i)/T_i}+1)$ as explicitly shown in the Supplemental Material~\cite{Supple}.

In Fig.~\ref{fig:Statresponse} we show the electrical and energy currents calculated from $G^<$ for three different values of $\Gamma$ and three values of $\alpha$ (solid lines) in comparision to the result obtained from the Luttinger and Einstein relation, i.e., by replacing the temperature and
chemical potential profile by the corresponding gravitational and electrical potentials (dashed lines). The plot shows that these quantities are unrelated and the discrepancies remain and become larger when $\Gamma$ or $\alpha$ is reduced. For vanishing phonon coupling, $\alpha=0$, and low $T$ one can calculate the heat currents induced by the couplings to the wire analytically~\cite{Supple}
\begin{align}
J^E_y\approx -\frac{\pi^2 k_B^2 T}{6h} 
\frac{\Gamma}{\pi M}\left (\partial_x T + \frac{v^2}{6 M^2} \partial_x^3 T \right).
\end{align}
It is non-universal and linear in $\Gamma$. Similarly, we find numerically that $J^E_y\propto \alpha$ for $\Gamma\to 0$ at low $T$, see Fig.~\ref{fig:Statresponse}. The  interpretation of this result is that some heat tunnels through the gapped topological insulator. This type of transport does, however, vanish for small $\Gamma$ and $\alpha$.  We conclude that spatially varying temperatures and chemical potentials do  not induce any universal topological currents in Chern insulators.


\begin{figure}[t]
\includegraphics[width=\columnwidth]{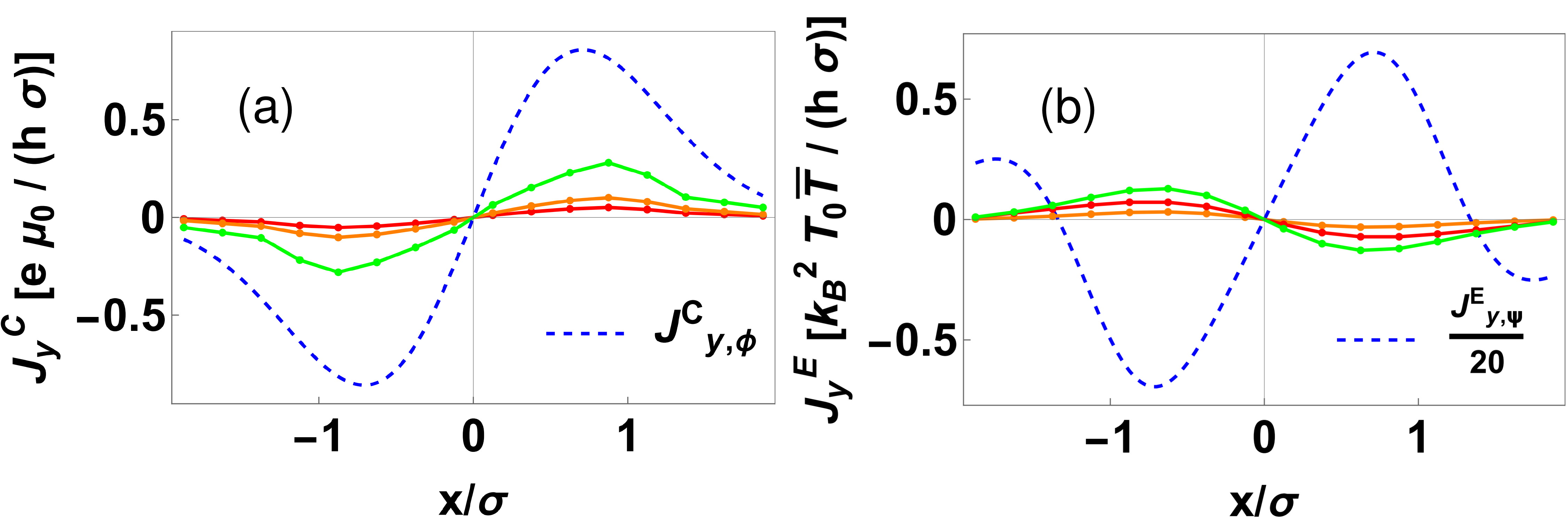}
\caption{{\bf Statistical response.} 
(a) The electrical Hall current $J_y$ in response to chemical potential $\mu (x) = \mu_0 \exp (-x^2/\sigma^2)$ and (b) the thermal Hall current $J_y^{E}$ in response to temperature $T (x) = \overline{T}+ T_0 \exp (-x^2/\sigma^2)$ calculated for topological phase of a square lattice model (see Supplemental Material for more details) with $v=\lambda=|M|$, $\overline{T} / |M| = 0.05$, $ T_0 /|M| = 0.0125$. The model includes both wires and phonon baths coupled to each lattice site. Parameters: $\Gamma/|M|= 0.05, 0.1, 1$, $\alpha=0.1$ bottom to top in panel a); $\Gamma/|M| = 0.01$, $\alpha = 0.1, 0.5, 1$  in panel b).
The blue dashed curves are drawn under the assumption that Einstein and Luttinger relations, Eq.~\eqref{eq:einsteinR} and Eqs.~\eqref{eq:analogy} and \eqref{eq:Haldanenonrelativisticresponse} are valid, showing that those relations {\em cannot} be used to describe this statistical response (the curve in (b) is drawn with the 20-fold reduced value for better visibility).}
\label{fig:Statresponse}
\end{figure}

While small space-dependent temperature profiles thus do not affect the gapped bulk, the situation is qualitatively different along the gapless chiral edge. The dissipationless heat current along the edge satisfies 
\begin{align}
\frac{ d J^E_{\text{edge}}(T)}{d T} =(c_R-c_L) \frac{\pi k_B^2}{6 \hbar} T. 
\end{align}
This implies immediately an anomaly-like source term on the right-hand side of the continuity equation for energy
\begin{align}
\frac{d}{dt} e_\text{edge} + \partial_x J^E_{\text{edge}}=  (c_R-c_L) \frac{\pi k_B^2}{6 \hbar} T \partial_x T, \label{eq:Tanomaly}
\end{align}
which is {\em linear} in the first derivative of temperature in contrast to the gravitational anomaly 
\begin{align} 
\frac{d}{dt} e_\text{edge} + \partial_x J^E_{\text{edge}} =  (c_R-c_L) \frac{\hbar }{24 \pi} v^2 \partial_x^3 \psi, \label{eq:Gravitanomaly}
\end{align}
where $v$ is the velocity of the edge, assumed here to be identical for all edge modes. In the supplemental material~\cite{Supple} we show that this formula remains valid in the non-relativistic setting, in contrast to Eq.~\eqref{eq:Haldanenonrelativisticresponse}. Note that the prefactor of the $\partial_x^3 \psi$ (Ref.~\cite{Stone2012} gets $48$ instead of $24$) depends on the used definition of $J^E_{\text{edge}}$ as discussed in the supplemental material~\cite{Supple}.
If one considers a stationary temperature profile induced, e.g., by the coupling to acoustic phonons, Eq.~\eqref{eq:Tanomaly} predicts the constant production
of energy.  As has been explored in detail in Refs.~\cite{Vinkler2018,Ye2018}, this flow of energy will, however, {\em not} result in some dissipationless bulk current but instead will go into the acoustic phonon system.
In Refs.~\cite{Vinkler2018,Ye2018} it has been shown that this effect is an essential prerequisit for the experimental observation of an approximately quantized thermal Hall effect.

{\em Conclusion.---}
In this paper we have analyzed and clarified to what extent the Luttinger relation, Eq.~\eqref{eq:luttingerR}, the Einstein relation,  Eq.~\eqref{eq:einsteinR},
   and the gravitational analogy, Eqs.~\eqref{eq:analogy} and \eqref{eq:relativisticresponse} can be used to describe the response to
 local and static variations of
temperature and chemical potential in Chern insulators. It turns out that none of these relations apply. A local temperature profile, 
   $T(\vec r)$, imprinted by heating a Chern insulator locally  does, for example, not produce any intrinsic heat currents at least as long as $T(\vec r)$
   remains small compared to the gap. Thus the physics of a space-dependent temperature is completely different from the physics of a gravitational potential which produces heat currents. Similarly, local variations of the  chemical potential, $\mu(\vec r)$, do not induce intrinsic transverse electric currents, while an external potential does. Ultimately, this difference can be traced back to the fact that space-dependent $T(\vec r)$ and $\mu(\vec r)$ simply do not show up as terms in the Hamiltonian but are effective quantities encoded in distribution functions which arise either from the coupling to a local bath or by local equilibration.
The Luttinger and Einstein relations remain fully valid only in the so-called transport limit ($\Omega \to 0$ after $\vec q \to 0$).

We have also shown that the gravitational bulk response of Chern insulators is not 
universal but depends on high-energy properties of the model. Transverse energy currents proportional to the third derivative of the gravitational potential may even be induced in topologically trivial phases. Unfortunately, such dissipationless energy currents are much more difficult to measure than the magnetic field created by their electric counterpart. It would be interesting to explore whether it is possible to induce gravitational potentials (e.g., by modulating the laser intensity) and measure such
currents in ultracold-atom experiments in optical lattices, using, e.g., time-of-flight measurements. 
\begin{acknowledgments}
\textit{Acknowledgments.---}
We thank Martin Zirnbauer and Ady Stern for useful discussions and Philipp Rosch for graphics support. Financial support of the Deutsche Forschungsgemeinschaft (DFG, German Research Foundation) within CRC1238 (project number 277146847, C02 and C04) 
and CRC183 (project number 277101999, A01 and A04) is acknowledged.
\end{acknowledgments}
%

\title{Supplemental Material for "Thermal Hall response: violation of gravitational analogues and Einstein relations"}
\author{Jinhong Park}
\affiliation{Institute for Theoretical Physics, University of Cologne, Z{\"u}lpicher Str. 77, 50937 K{\"o}ln, Germany}
\author{Omri Golan}
\affiliation{Department of Condensed Matter Physics, Weizmann Institute of Science, Rehovot 76100, Israel}
\author{Yuval Vinkler-Aviv} 
\affiliation{Institute for Theoretical Physics, University of Cologne, Z{\"u}lpicher Str. 77, 50937 K{\"o}ln, Germany}
\author{Achim Rosch}
\affiliation{Institute for Theoretical Physics, University of Cologne, Z{\"u}lpicher Str. 77, 50937 K{\"o}ln, Germany}
\date{\today}
\maketitle

\section{Response to electrostatic or gravitational potential}
In this section, we analytically calculate the Hall response to a smoothly varying electrostatic potential $\phi (\vec{r})$ and gravitational potential $\psi (\vec{r})$ at zero chemical potential and temperature.

\label{Appen:mechanicalresponse}
\subsection{Response to electrostatic potential bump}

The calculation is performed in the continuum limit of the Haldane model (Eq.~(9) in the main text). In the momentum space, the system is described by the Hamiltonian
\begin{align} \label{systemHamiltonian}
\hat{H}_0 &= \sum_{\vec{k}} \Psi^{\dagger} (\vec{k}) \left [ v \vec{k} \cdot \vec{\sigma} + (M + \lambda^2 \vec{k}^2) \sigma_z \right ]\Psi (\vec{k})
\nonumber \\ &= \sum_{\vec{k}} \Psi^{\dagger} (\vec{k}) \hat{h} \left [\vec{k} \right]\Psi (\vec{k})
\end{align}
with a vector of the Pauli matrices  $\vec{\sigma} = (\sigma_x, \sigma_y)$. The two component spinor $\Psi^{\dagger} (\vec{k}) = (\Psi_{1}^{\dagger} (\vec{k}), \Psi_{2}^{\dagger} (\vec{k}))$ creates electrons with momentum $\vec k = (k_x, k_y)$. A static electrostatic potential is linearly coupled to the system as 
\begin{align}
\hat{H}_\phi &=  \sum_{\sigma = 1, 2} \int d\vec{r} \phi(\vec{r})\Psi_{\sigma}^{\dagger} (\vec{r}) \Psi_{\sigma} (\vec{r}).  
\end{align}
For simplicity, the electrostatic potential is assumed to vary in the $x$ direction while being constant in the $y$ direction.

The continuity equation for the charge density results in
\begin{align} \label{supeq:continuityEqcharge}
\nabla \cdot \hat{\vec{J}}{}^{C} (\vec{r}, t) &= e \frac{d \hat{n} (\vec{r}, t)}{dt} = \frac{i e}{\hbar} \left [\hat{H}_0, \hat{n} (\vec{r}, t) \right ]
 \nonumber \\  & = - \frac{ie}{\hbar} \int \frac{d \vec{K}}{(2\pi)^2} \sum_{\vec{q}} 
e^{i \vec{q} \cdot \vec{r}} \Psi^{\dagger} \left (\vec{K} - \frac{\vec{q}}{2}, t \right ) 
\nonumber \\ & \times
\left ( v \vec{q} \cdot \vec{\sigma}  + 2 \vec{K}\cdot \vec{q} \lambda^2 \sigma_z \right ) \Psi \left (\vec{K} + \frac{\vec{q}}{2}, t \right ). 
\end{align}
Here $-e$ is the electron charge with $e>0$. 
From Eq.~\eqref{supeq:continuityEqcharge}, the electrical current density operator can be identified as 
\begin{align} \label{supeq:currentdensityop}
\hat{\vec{J}}{}^{C} (\vec{r}) = - & \frac{e}{\hbar}\int \frac{d \vec{K}}{(2\pi)^2} \sum_{\vec{q}} 
e^{i \vec{q} \cdot \vec{r}} \Psi^{\dagger} \left (\vec{K} - \frac{\vec{q}}{2} \right ) \nonumber \\ 
& \times
\left ( v \vec{\sigma} + 2 \vec{K} \lambda^2 \sigma_z \right ) \Psi \left (\vec{K} + \frac{\vec{q}}{2} \right ). 
\end{align}
We consider the expectation value $\vec{J}^{C}_{\phi} (\vec{r}) \equiv \langle \hat{\vec{J}}{}^{C} (\vec{r})  \rangle $, written as 
\begin{align} \label{currentexpvalue}
\vec{J}_{\phi}^{C} (\vec{r}) = & \frac{ie}{\hbar}  \int \frac{d \vec{K}}{(2\pi)^2}\sum_{ \vec{q}} \int \frac{d \omega}{ 2 \pi}
\textrm{Tr} \Big [ 
\left ( v \bm{\sigma}  + 2 \vec{K} \lambda^2 \sigma_z \right ) \nonumber \\ 
& \times
G^{<}_{\vec{K} + \vec{q}/2,  \vec{K}- \vec{q}/{2}} \left (\omega \right ) \Big ] , 
\end{align}
in the presence of a static electrostatic potential bump [$\phi (\vec{r})$]. 
The average is taken over the 
the eigenstates of the Hamiltonian $H_0 + H _{\phi}$. The trace is performed over the bands of the continuum model. 
The lesser Green's function $G^<$ can be obtained by expanding 
up to the first order in $\phi (\vec{r})$ as
\begin{align} \label{supeq:langrethrule}
G_{\sigma' \vec{k'}, \sigma \vec{k}}^{<} (\omega) = &  g_{\sigma' \vec{k}, \sigma \vec{k}}^{<} (\omega) \delta_{\vec{k} \vec{k'}} 
+ \sum_{\vec{r}_1} \sum_{\sigma_1} e^{i (\vec{k}- \vec{k'})\cdot \vec{r_1}} \phi (\vec{r_1}) \nonumber \\ 
\times &  
\Big ( g_{\sigma' \vec{k'}, \sigma_1 \vec{k'}}^{R} (\omega)  g_{\sigma_1 \vec{k}, \sigma \vec{k}}^{<} (\omega)
\nonumber \\ 
+ & g_{\sigma' \vec{k'}, \sigma_1 \vec{k'}}^{<} (\omega)  g_{\sigma_1 \vec{k}, \sigma \vec{k}}^{A} (\omega)
 \Big ).
\end{align}
The $g$'s are the Green's functions in the absence of the electrostatic potential, explicitly written as 
\begin{align} \label{supeq:retadlessergreenfun}
g^{R/A}_{ \vec{k'},  \vec{k}}  (\omega) &=\frac{\delta_{\vec{k}, \vec{k'}}}{ (\omega - h[\vec{k}]\pm i\eta)}, \nonumber \\  
g^{<}_{\vec{k'}, \vec{k}} (\omega) & = f_0 (\omega) \delta_{\vec{k}, \vec{k'}} 
\left (\frac{1}{\omega - h[\vec{k}]- i \eta}-
\frac{1}{\omega - h[\vec{k}]+ i \eta} \right).
\end{align}
Plugging Eq.~\eqref{supeq:langrethrule} into 
Eq.~\eqref{currentexpvalue}, we obtain the charge current density $J_{y, \phi}^{C} ( \vec{r})$ flowing along the $y$ direction (the Hall response)
\begin{align} \label{supeq:HallcurrentMechForce}
J_{y, \phi}^{C} ( \vec{r}) =& 2\frac{e}{\hbar}\int \frac{d\omega}{2\pi} \int \frac{d\vec{K}}{(2\pi)^2} \int \frac{d\vec{q}}{(2\pi)^2} 
 \int d\vec{r_1} \phi (\vec{r}_1)  f_0 (\omega) \nonumber \\ & \times
  \textrm{Im} \bigg [e^{i \vec{q} \cdot (\vec{r} - \vec{r}_1)} \textrm{Tr}\bigg [ \left ( v \sigma_y 
+ 2 \lambda^2 \sigma_z K_y \right )  \nonumber \\ 
& \times
\frac{1}{\omega- h (\vec{K} +\vec{q}/2)  + i \eta } \frac{1}{\omega- h (\vec{K} -\vec{q}/2)  + i \eta } \bigg ] \bigg ].
\end{align}
%
Being interested in the long-range physics, we expand the term inside the trace in momentum $\vec{q}$. 
The leading contribution comes from the term linear in $q_x$. The direct calculation of the integral of the linear term $\sim q_x$ in Eq.~\eqref{supeq:HallcurrentMechForce} results in 
\begin{align} \label{analyticMechanicalforce}
J_{y, \phi}^{C} (\vec{r}) \approx& 2 \frac{e}{\hbar} \int_{-\infty}^{0} \frac{d \omega}{2\pi} \int \frac{d K}{2\pi}  \partial_x \phi(\vec{r}) 
\nonumber \\ & \times
 \textrm{Im} \left [\frac{2 v^2 K (\lambda^2 K^2 - M)}
{\left ((\lambda^2 K^2 + M)^2 + K^2 v^2 + (\eta - i \omega)^2 \right)^2} \right ] 
\nonumber \\ =& 
-\frac{e \partial_x \phi(\vec{r})}{2\pi \hbar} \theta(-M).
\end{align}
 The transverse conductivity is fully universal and given by $\sigma_{xy} = e / (2\pi \hbar)$ ($\sigma_{xy} = 0$) in the topological (trivial) phase [$M<0$ ($M>0$)].

\subsection{Gravitational response}
\label{appsubsection:graviationalresponse}
We next turn our intention to the thermal response to a gravitational potential bump $\psi (\vec r)$. 
The gravitational potential is linearly coupled to the Hamiltonian density $\hat{h}_{c} (\vec{r})$ of the continuum limit of the Haldane model as
\begin{align}
\hat{H}_{\psi} &= \int d^2 \vec{r} \left ( 1 + \psi (\vec {r}) \right ) \hat{h}_c (\vec r). 
\end{align}
In the momentum space, the Hamiltonian density $\hat{h}_c (\vec{r})$ reads
\begin{align} \label{supeq:Hamiltoniandensity}
  \hat{h}_c (\vec{r})  = \int \frac{d\vec{k}}{(2\pi)^2}\sum_{ \vec{k'}} e^{i (\vec{k'} -  \vec{k}) \cdot \vec{r}} 
\Psi^{\dagger}_{\vec{k}} \left (\vec{d} (\vec{k}, \vec{k'}) \cdot \vec{\sigma} \right)  \Psi_{\vec{k'}} ,
\end{align}
with \begin{align} \label{supeq:haldaneenergy}
\vec{d} (\vec{k}, \vec{k'}) = \hbar v \left (\frac{\vec{k} + \vec{k'}}{2} \right) + \hbar \left (M + \lambda^2 \vec{k} \cdot \vec{k'} \right ) \hat{\vec{z}}.
\end{align}
The energy current operator $\hat{\vec{J}}{}_{\psi}^{E}$ can be derived from the continuity equation for the energy density
\begin{align} \label{supeq:continuityEnergy}
\nabla \cdot \hat{\vec{J}}{}_{\psi}^{E} (\vec r)  &= \frac{ i}{\hbar} (1 + \psi (\vec {r}) ) [\hat{h}_c (\vec r), \hat{H}_{\psi}]
\nonumber \\ & = \frac{i}{\hbar} (1 + \psi (\vec {r}) ) \int d^2\vec{r'} (1 + \psi (\vec {r'}) ) [\hat{h}_c (\vec r), \hat{h}_c (\vec r')].
\end{align}  
Inserting Eq.~\eqref{supeq:Hamiltoniandensity} into Eq.~\eqref{supeq:continuityEnergy} and employing the anti-commutation relation for the fermion fields, we arrive 
\begin{widetext}
\begin{align}
\nabla \cdot \hat{\vec{J}}{}_{\psi}^{E} (\vec r) &= \frac{i}{\hbar} (1 + \psi (\vec {r}) ) \int \frac{d \vec{k}''}{(2 \pi)^2}
\sum_{\vec{k}, \vec{k'}}\int d^2 \vec{r'} (1 + \psi (\vec {r'}) ) 
\nonumber \\ & \times
\Psi^{\dagger}_{\vec{k}}   \bigg \{  \left ( e^{i 
\left ((\vec{k''} - \vec{k}) \cdot \vec{r} - (\vec{k''} - \vec{k'}) \cdot \vec{r'} \right )} - e^{i 
\left ( (\vec{k''} - \vec{k}) \cdot \vec{r'} - (\vec{k''} - \vec{k'}) \cdot \vec{r}\right ) }  \right )  \left (d (\vec{k}, \vec{k''}) \cdot  d (\vec{k''}, \vec{k'}) + i  \sigma \cdot  d (\vec{k}, \vec{k''}) \times  d (\vec{k''}, \vec{k'}) \right )  \bigg \}  \Psi_{\vec{k'}}, 
\end{align}
We (i) replace $\vec{k''}$ with the derivative with respect to $\vec{r'}$
\begin{align}
\vec{k''}e^{i 
\left ((\vec{k''} - \vec{k}) \cdot \vec{r} - (\vec{k''} - \vec{k'}) \cdot \vec{r'} \right )}
&= \left (i \nabla_{\vec{r'}} + \vec{k'} \right ) e^{i 
\left ((\vec{k''} - \vec{k}) \cdot \vec{r} - (\vec{k''} - \vec{k'}) \cdot \vec{r'} \right )} \nonumber \\ 
\vec{k''}e^{i 
\left ( (\vec{k''} - \vec{k}) \cdot \vec{r'} - (\vec{k''} - \vec{k'}) \cdot \vec{r}\right ) }
&= \left (-i \nabla_{\vec{r'}} + \vec{k} \right )e^{i 
\left ( (\vec{k''} - \vec{k}) \cdot \vec{r'} - (\vec{k''} - \vec{k'}) \cdot \vec{r}\right ) }, 
\end{align}
(ii) integrate over $\vec{k''}$ to obtain the delta function $\delta ( \vec{r} - \vec{r'})$, and 
(iii) use the integration by parts to move the derivative to act on $(1 + \psi (\vec{r'}) )$. Those procedures (i), (ii), (iii) result in  
\begin{align} \label{Divenergycurrent}
\nabla \cdot \hat{\vec{J}}{}_{\psi}^{E} (\vec r) &= \frac{i}{\hbar} (1 + \psi (\vec {r}) ) \int \frac{d \vec{k}}{(2 \pi)^2}
\sum_{ \vec{k'}} e^{i (\vec{k'} - \vec{k}) \cdot \vec{r}}  \Psi^{\dagger}_{\vec{k}}
[ d (\vec{k}, \vec{k'} - i \nabla_{\vec{r}}) \cdot d (\vec{k'} - i \nabla_{\vec{r}}, \vec{k'}) - 
d (\vec{k}, \vec{k} + i \nabla_{\vec{r}}) \cdot 
d (\vec{k} +  i \nabla_{\vec{r}}, \vec{k'})
  \nonumber \\ & +
 i  \sigma \cdot  [ d (\vec{k}, \vec{k'} - i \nabla_{\vec{r}}) \times d (\vec{k'} - i \nabla_{\vec{r}}, \vec{k'}) -
d (\vec{k}, \vec{k} + i \nabla_{\vec{r}}) \times
d (\vec{k} + i \nabla_{\vec{r}}, \vec{k'})] (1 + \psi(\vec{r}))  \Psi_{\vec{k'}}.
\end{align} 
The right hand side of Eq.~\eqref{Divenergycurrent} has a form of $\nabla \cdot \left [ (1 + \psi (\vec{r}))^2 
\hat{\vec{J}}{}^{E} (\vec{r}) \right] $, where the energy current density $\hat{\vec{J}}{}^{E} (\vec{r})$ in the absence of the gravitational potential is written as
\begin{align} \label{energycurrentabs}
\hat{\vec{J}}{}^{E} (\vec{r})  &= \hbar \int\frac{d \vec{K}}{ (2 \pi)^2} \sum_{ \vec{q}} \Psi^{\dagger}_{\vec{K}-\vec{q}/2}
 e^{i \vec{q} \cdot \vec{r}} \vec{J}^E_{\vec{K}- \vec{q}/2; \vec{K} +\vec{q}/2} \Psi_{\vec{K}+\vec{q}/2}, \nonumber \\ \vec{J}^{E}_{\vec{K} - \vec{q}/2; \vec{K} + \vec{q}/2} & = 
 \left ( v^{2} + 2 \lambda^2 \left (M + \lambda^2 \vec{K}^2  \right ) \right) \vec{K} + 
 \lambda^2  \left (i v \vec{\sigma} \cdot \left (\vec{K} \times \hat{z} \right)- \frac{\lambda^2}{2} \vec{K} \cdot \vec{q}\right) \vec{q} 
 - \frac{i}{4} \left (v^2\sigma_z - 2  v \lambda^2 \sigma \cdot \vec{K} \right)\left ( \hat{\vec{z}} \times \vec{q} \right ) .
\end{align}
Here, we have used central mass momentum $\vec{K} = \left (\vec{k} + \vec{k'} \right )/2$ and relative momentum $\vec{q} = \vec{k'} - \vec{k}$. Note that the locality condition~\cite{Bradlyn2015, Cooper1997, Qin2011} is fulfilled such that the energy current is uniquely defined as $\hat{\vec{J}}{}_{\psi}^{E} (\vec r) = (1 + \psi (\vec{r}))^2 \hat{\vec{J}}{}^{E} (\vec{r})$. 

We next consider the expectation value $\vec{J}^{E}_{
\psi} (\vec r) \equiv \langle\hat{\vec{J}}{}_{
\psi}^{E} (\vec r)  \rangle  $ in the presence of the gravitational field $\psi (\vec{r})$, given by 
\begin{align} \label{supeq:energycurrentexp}
\vec{J}^{E}_{\psi} (\vec r) &= - i \hbar \left (1 + \psi (\vec{r}) \right)^2 \sum_{\vec{q}} \int \frac{d \vec{K}}{(2\pi)^2}  \int \frac{d \omega}{2\pi} e^{i \vec{q} \cdot \vec{r}} \nonumber \\ 
& \times \vec{J}^E_{\vec{K} - \vec{q}/2, \sigma; \vec{K} + \vec{q}/2, \sigma'} G_{\vec{K} + \vec{q}/2, \sigma'; \vec{K} - \vec{q}/2, \sigma}^{<} (\omega). 
\end{align}
The average is taken over the 
the eigenstates of the Hamiltonian $H _{\psi}$.
The lesser Green's function $G^<$ can be obtained from the standard linear response theory, expanded 
up to the first order in the static $\psi (\vec{r})$

\begin{align} \label{lessGreengravfield}
G_{\sigma,\vec{k}; \sigma', \vec{k'}}^{<} (\omega) & =   g_{\sigma \sigma' }^{<} (\vec{k},\omega) \delta_{\vec{k} \vec{k'}} 
+ \sum_{\vec{r_1}} e^{i (\vec{k'}- \vec{k})\cdot \vec{r_1}} \psi (\vec{r_1})  
\sum_{\sigma_1, \sigma_2}\Big ( g_{\sigma \sigma_1}^{R} (\vec{k}, \omega) \left( d (\vec{k}, \vec{k'}) \cdot \sigma \right)_{\sigma_1 \sigma_2} g_{\sigma_2 \sigma' }^{<} (\vec{k'}, \omega) \nonumber \\ &
+ g_{\sigma  \sigma_1}^{<} (\vec{k}, \omega) \left( d (\vec{k}, \vec{k'}) \cdot \sigma \right)_{\sigma_1 \sigma_2}g_{\sigma_2 \sigma'}^{A} ( \vec{k'}, \omega)
 \Big ).
\end{align}
Here the $g$'s denote Green's functions in the absence of $\psi (\vec{r})$ as explicitly written in Eq.~\eqref{supeq:retadlessergreenfun}. 
Plugging Eq.~\eqref{lessGreengravfield} into Eq.~\eqref{supeq:energycurrentexp} results in 
\begin{align} \label{ThermalcurrentGravit}
\vec{J}^{E}_{\psi} (\vec r)=& - 2 \left (1 + \psi (\vec{r}) \right)^2  \int d\vec{r_1} \int \frac{d^2 \vec{K}}{(2\pi)^2} \int \frac{d^2 \vec{q}}{(2\pi)^2}\int \frac{d \omega}{2\pi}   f_0 (\omega)\psi (\vec{r_1}) 
\textrm{Im} \Big[ e^{i \vec{q} \cdot (\vec{r} -\vec{r_1})}
\textrm{Tr} \Big [\vec{J}^{E}_{\vec{K} - \vec{q}/2; \vec{K} + \vec{q}/2} \nonumber \\ \times &  g^R ( \vec{K} + \vec{q}/2, \omega) 
 h_{ \vec{K} + \vec{q}/2; \vec{K} - \vec{q}/2}
g^R ( \vec{K} - \vec{q}/2, \omega) \Big] \Big ]. 
\end{align}
Here $h_{0,\vec{k};\vec{k'}} = \hbar v\vec{\sigma} \cdot \left ( \vec{k}+ \vec{k'}\right )/2 + \hbar \sigma_z \lambda^2 \left (\vec{k} \cdot \vec{k'} \right)$ arises from the Fourier transformation of the energy density.
Being interested in the long-range physics, we expand the term inside the trace in momentum $q_x$. 
The leading contribution comes from the qubic term $q_x^3$. Performing the direct calculation of the integral only for the qubic term, we obtain the energy current $J_{y, \psi}^{E} (\vec r)$, flowing along the $y$ direction

\begin{align}
J_{y, \psi}^{E} (\vec r) = - \frac{3 \hbar}{96}\left (1 + \psi (\vec{r}) \right)^2 \partial_x^3  \psi (\vec{r})  \int \frac{d K}{(2 \pi)} \frac{K^3 (\lambda^2 K^2 - M) (2 \lambda^4 K^2 + 2 \lambda^2 M + v^2)^2}{ (\lambda^4 K^4 + 2 \lambda^2 K^2 M + M^2 + K^2 v^2)^2}. 
\end{align}
The integral can be further performed with the momentum cutoff $K_{\rm{cut}}$, resulting in 
\begin{align} \label{supeq:nonrelativisticresponse}
J_{y, \psi}^{E} (\vec r) &= \frac{\hbar}{96 \pi} 
\left (1 + \psi (\vec{r}) \right)^2 \partial_x^3  \psi (\vec{r}) \Bigg (
(-2 \lambda^2 M + v^2) \textrm{sgn} (M) + (2 \lambda^2 M + 5 v^2) \nonumber \\ 
& - 3 v^2 
\log \left (\frac{v^2 + 2 K_{\rm{cut}}^2 \lambda^4+ 2\lambda^2 (M+ \sqrt{M^2+ (2 \lambda^2 M + v^2)  K_{\rm{cut}}^2 + \lambda^4 K_{\rm{cut}}^4})}{v^2 + 2 \lambda^2 (M + |M|)} \Bigg )
\right ). 
\end{align}
In large momentum cutoff limit, Eq.~\eqref{supeq:nonrelativisticresponse} is approximated as Eq.~(10) in the main text. 
If one repeats the same calculation sketched above in the relativistic case of $\lambda = 0$ and $v \rightarrow c$,  one obtains instead for the thermal current flowing along the $y$ direction
\begin{align} \label{supeq:relativisticresponse}
J_{y, \psi}^{E} (\vec r) = - \frac{\hbar c^2}{96 \pi} (1 + \psi (\vec{r}))^2 \partial_x^3  \psi (\vec{r}) \textrm{sgn} (-M). 
\end{align}
Note that the relativistic response is {\em not} simply obtained by taking the limit $\lambda \to 0$, $K_{\rm cut} \to \infty$ of the non-relativistic model.
This relativistic topological gravitational response, Eq.~\eqref{supeq:relativisticresponse}, coincides with the response derived from the gravitational Chern-Simon term in Ref.~\cite{Stone2012}.
\end{widetext}

 

\section{One dimensional gravitational anomaly} \label{Appendix:1danomaly}

In this section, we consider the gravitational anomaly in an one-dimensional edge channel. We show that the anomaly term
does not depend on the full dispersion of the edge mode, but only on the Fermi edge velocity.


We consider an one-dimensional edge channel propagating along the $x$ direction, coupled to a gravitational potential $\psi (x)$ as
\begin{align}
    \hat{H}_{\psi} = \int dx (1 + \psi (x)) \hat{h}_{\rm{edge}} (x). 
\end{align}
The edge channel is assumed to have both linear and quadratic dispersion around Fermi momenta $k = 0$ as
\begin{align}
    \hat{h}_{\rm{edge}} (x) & = \int \frac{dk}{(2\pi)}\sum_{k'} e^{i (k' -  k) x} 
\Psi^{\dagger}_{k} h_{ k;k'}  \Psi_{k'} \nonumber
\end{align}
with 
\begin{align}
h_{ k;k'} = \hbar v_1 \left (\frac{k + k'}{2} \right) + 
\hbar v_2 k k'. 
\end{align}
Employing the same procedure as specified in Eqs.~\eqref{supeq:continuityEnergy}-\eqref{energycurrentabs} of App.~\ref{appsubsection:graviationalresponse}, we find the energy current for this one-dimensional model
\begin{align}
J^{E}_{ k;k'}= h_{ k;k'} \left(v_1 + v_2 (k + k') \right). 
\end{align}
The 1d energy current $J^{E}_{\textrm{edge}} (x)$ in the presence of the gravitational potential 
is given by 
\begin{align}
J^{E}_{\textrm{edge}} (x)=& - 2 \left (1 + \psi (x) \right)^2  \int dx_1 \int_{-\infty}^{\infty} \frac{d K}{(2\pi)} \int_{-\infty}^{\infty} \frac{d q}{(2\pi)}
\nonumber \\ 
\times & \int \frac{d \omega}{2\pi}   f_0 (\omega)\psi (x_1) 
\textrm{Im} \Big[ e^{i q (x - x_1)}
J^E_{K - q/2; K + q/2} \nonumber \\ \times &  g^R ( K + q/2, \omega) 
 h_{ K+q/2;  K-q/2}
g^R ( K-q/2, \omega) \Big ]. 
\end{align}
We next use the gradient expansion approach discussed in App.~\ref{Appen:mechanicalresponse}. The leading contribution comes from the quadratic term in $q$. Direct integration  only for the quadratic term over $q$, $K$ and $x_1$ results in 
\begin{align}
    J^{E}_{\textrm{edge}} (x) = \frac{\hbar}{24 \pi} 
   \textrm{sgn} (v_1) v_1^2 \partial_x^2 \psi (x). 
\end{align}
The corresponding gravitational anomaly equation reads 
\begin{align} \label{supeq:1danomaly}
\frac{d}{dt} e_{\textrm{edge}} + \partial_x J^{E}_{\textrm{edge}} &= \frac{\hbar}{24 \pi}    \textrm{sgn} (v_1)
    v_1^2 \partial_x^3\psi (x) \nonumber \\ 
     &= \frac{\hbar}{24 \pi}    (c_R - c_L)
    v_1^2 \partial_x^3\psi (x). 
\end{align}
Here, we added the time derivative of the energy density $\frac{d}{dt} e_{\textrm{edge}}$ by hand, which is zero in the static perturbation. Furthermore, we generalize our results to the case with several chiral modes in the second equality, where the velocities of the modes are assumed to be identical. 
Importantly, there is no $v_2$ dependence on Eq.~\eqref{supeq:1danomaly}. Eq.~\eqref{supeq:1danomaly} shows that the 1d gravitational anomaly depends on the Fermi velocity rather than the entire dispersion of the edge channel. 

In Ref.~\cite{Stone2012} a similar equation is obtained in the relativistic limit with  $v_1$ replaced by the speed of light (set to $1$ in Ref.~\cite{Stone2012}) and $c_R-c_L$ denoted by $c$.
Furthermore, the result quoted by Stone, Eq.~(75) of Ref.~\cite{Stone2012}, is a factor of $2$ smaller. This discrepancy can be traced back
to a different definition of the heat current. We recover the result of Stone by redefining the energy-momentum tensor $T^{\mu \sigma} \to T^{\mu \sigma}+c^2 \frac{c_R-c_L}{96 \pi}\frac{1}{\sqrt{g}} \epsilon^{\mu \sigma} R$. This transformation ensures that $T^{\mu \sigma}=T^{\sigma \mu}$. 

In the relativistic case bulk and edge anomalies match. To see this, consider a spatially varying mass gap in the $y$ direction with $M(y) > 0$ ($M(y) < 0$) for $y > 0$ ($y <0$).  In this case the relativistic theory, Eq.~\eqref{supeq:relativisticresponse}, predicts that an energy current is flowing towards the boundary, which precisely matches the edge anomaly $\frac{\hbar}{48 \pi}    (c_R - c_L)
    c^2 \partial_x^3\psi (x)$ for $c_R-c_L=1$ when using the symmetric version of $T^{\mu \sigma}$ discussed above. This is, however, not the case in the non-relativistic case, where the bulk response is non-universal and even the edge response depends on the non-universal Fermi velocity of the edge mode. The disagreement between Eq.~\eqref{supeq:nonrelativisticresponse} and Eq.~\eqref{supeq:1danomaly} implies that extra non-universal energy currents exist with components parallel to the boundary (i.e., along the $x$ direction) which are not described by the edge theory alone. 
The calculation of these extra edge currents is beyond the scope of this work. We expect that they arise from the scattering of bulk modes from the edge.

\section{Lattice calculation} \label{app:Latticecalculation}

In this section, we discuss the Haldane lattice model coupled to a smoothly varying gravitational (electrostatic) potential, and calculate the thermal (electrical) Hall response to the potential. 

\begin{figure} 
\includegraphics[width=0.9\columnwidth]{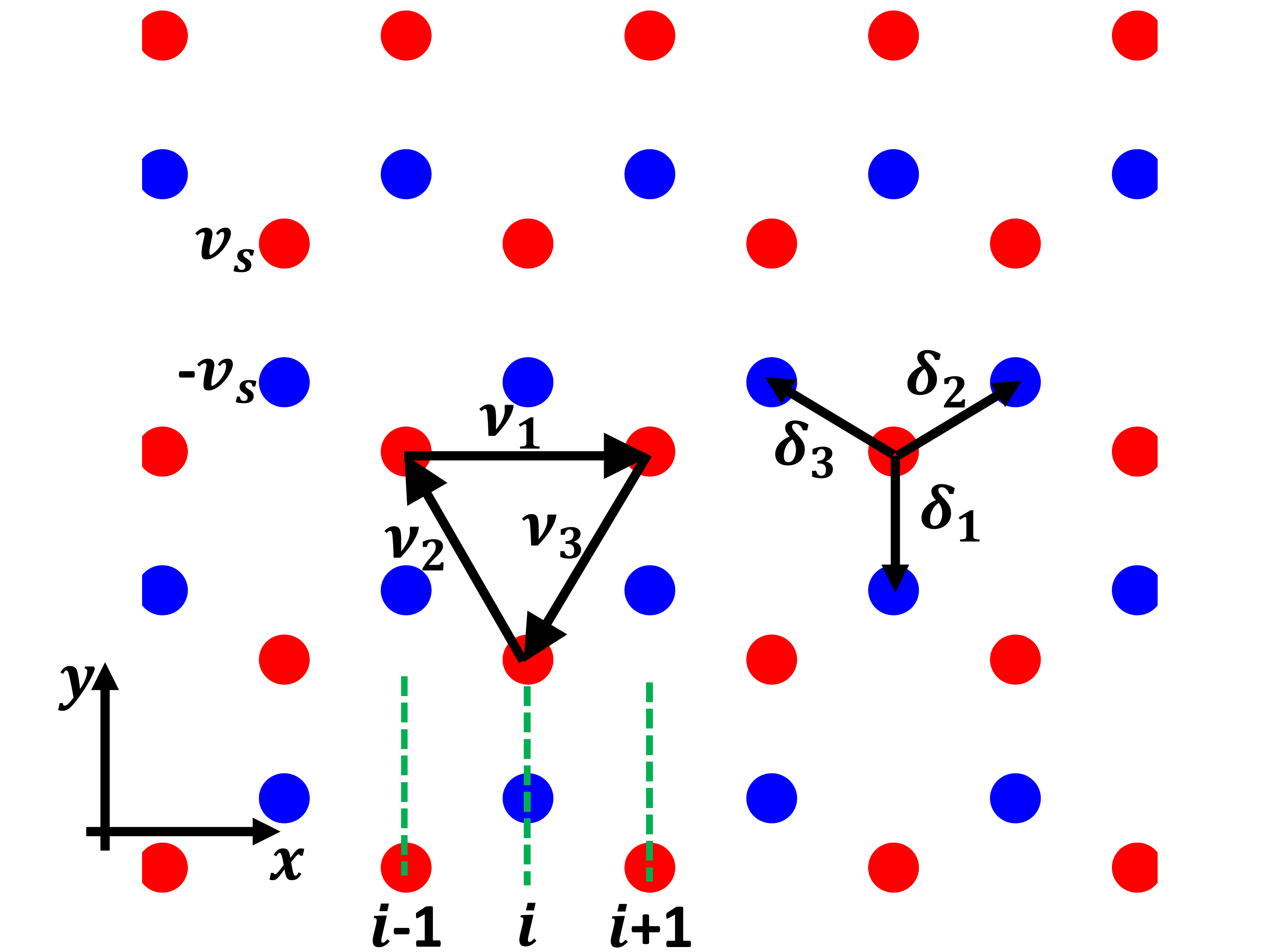} 
\caption{{\bf Haldane lattice model.} The honeycomb lattice with the nearest neighbor hoppings (right), the next-nearest neighbor hoppings with strength $-i t_2$ (middle) and stagger potentials $\pm v_s$ (right). For a lattice simulation, we consider a strip with finite size in the $x$ direction, but infinitely long along the $y$ direction.
}
\label{Fig:suppleHaldanemodel}
\end{figure}

We consider a honeycomb lattice (cf. Figs.~2 in the main text and \ref{Fig:suppleHaldanemodel}) that contains two sublattices, $A$ (denoted as red dots) and $B$ (denoted as blue dots) per unit cell. The lattice vectors are given by
\begin{align}
    \vec{a}_1 = a \left( \frac{1}{2},\frac{\sqrt{3}}{2} \right),\,\,\,\,\,\,
    \vec{a}_2 =  a \left( \frac{1}{2}, -\frac{\sqrt{3}}{2} \right),
\end{align}
with the lattice constant $a$, and the lattice sites that belongs to $A$ ($B$) sublattice can be written as
$\vec{r}_{i,A} = m_{1,i} \vec{a}_1 + m_{2,i} \vec{a}_2$ ($\vec{r}_{i,B} = m_{1,i} \vec{a}_1 + m_{2,i} \vec{a}_2 + \vec{\delta}_1$) with integers $m_{1,i}$ and $m_{2,i}$ (cf. Fig.~\ref{Fig:suppleHaldanemodel}).
The Haldane model is described by three terms: real-valued nearest-neighbor hoppings $-t_1$, purely imaginary next-nearest neighbor hoppings $\pm i t_2$, and a staggered potential which takes different values $\pm v_s$ on sublattices $A$ and $B$. 
\begin{align}
    \hat{H} =& \sum_{\vec{r}_i} \hat{h}_{\vec{r}_i}= \sum_{\vec{r}_i} \left (\hat{h}_{\rm{NN},\vec{r}_i}+\hat{h}_{\rm{NNN},\vec{r}_i} +\hat{h}_{\rm{sp},\vec{r}_i} \right),
\end{align}
where the local Hamiltonians are written as
\begin{align} \label{eq:supplelatticeHamiltonian} 
\hat{h}_{\rm{NN}, \vec{r}_i} = &- t_1  \sum_{\ell = 1,2,3} c_{\vec{r}_i, A}^{\dagger} c_{\vec{r}_i + \vec{\delta}_\ell - \vec{\delta}_1, B} + \textrm{H.c.}, \nonumber \\ 
\hat{h}_{\rm{NNN}, \vec{r}_i} = &-  i t_2 \sum_{\ell = 1,2,3}
\Big(c^{\dagger}_{\vec{r}_i + \vec{\sigma}_\ell, A} c_{\vec{r}_i, A} 
- c^{\dagger}_{\vec{r}_i  + \vec{\sigma}_\ell , B} c_{\vec{r}_i, B} \nonumber \\ & - \textrm{H.c.} \Big) \nonumber \\ 
\hat{h}_{\rm{sp}, \vec{r}_i} =& v_s \left (c_{\vec{r}_i,A}^{\dagger}c_{\vec{r}_i,A}-c_{\vec{r}_i,B}^{\dagger}c_{\vec{r}_i,B}\right).
\end{align}
Here $\vec{\delta}_1 = a (0, -1/ \sqrt{3} )$, $\vec{\delta}_2 = a (1, 1/ \sqrt{3} ) /2$,  $\vec{\delta}_3 = a (-1, 1/ \sqrt{3} )/2$, and $\vec{\sigma}_1 = a (1, 0)$, $\vec{\sigma}_2 = a (-1 , \sqrt{3})/2$, $\vec{\sigma}_3 = -a (1 , \sqrt{3})/2$ as shown in Fig.~\ref{Fig:suppleHaldanemodel}.

We next consider a smoothly varying gravitational potential $\psi (\vec{r})$ (electrostatic potential $\phi (\vec{r})$), locally coupled to the lattice Hamiltonian, Eq.~\eqref{eq:supplelatticeHamiltonian} (to the charge density). While the electrostatic potential is put on each lattice site, 
the gravitational potential is put on the center $(\vec{r}_i + \vec{r}_j)/2$ of the links by the Hamiltonian to connect lattice sites $\vec{r}_i$ and $\vec{r}_j$ as 
\begin{align}
    \hat{H} [\psi, \phi]  = \sum_{\vec{r}_i} \hat{h}_{\vec{r}_i} = & \sum_{\vec{r}_i} \Big (\hat{h}_{\rm{NN}, \vec{r}_i} [\psi]+\hat{h}_{\rm{NNN}, \vec{r}_i} [\psi] \nonumber \\ &+\hat{h}_{\rm{sp},\vec{r}_i} [\psi] + \hat{h}_{\rm{ep},\vec{r}_i} [\psi, \phi] \Big ),
\end{align}
with the local Hamiltonian
\begin{align} \label{eq:supplelatHamiltoniangrav}
\hat{h}_{\rm{NN}, \vec{r}_i} [\psi] = &- t_1  \sum_{\ell = 1,2,3} \left (1 + \psi (\vec{r}_i + \vec{\delta}_\ell/2) \right)
\nonumber \\ &\times
\left (c_{\vec{r}_i, A}^{\dagger} c_{\vec{r}_i + \vec{\delta}_\ell -\vec{\delta}_1, B} + \textrm{H.c.} \right), \nonumber \\ 
\hat{h}_{\rm{NNN}, \vec{r}_i} [\psi] = & -  i t_2 \sum_{\ell= 1,2,3}
\Big( (1 + \psi (\vec{r}_i + \vec{\sigma}_\ell/2)) c^{\dagger}_{\vec{r}_i + \vec{\sigma}_\ell, A} c_{\vec{r}_i, A} \nonumber \\ & - (1 + \psi (\vec{r}_i + \vec{\delta}_1+\vec{\sigma}_\ell/2)) c^{\dagger}_{\vec{r}_i  + \vec{\sigma}_\ell , B} c_{\vec{r}_i, B} \nonumber
\\ &
- \textrm{H.c.} \Big) \nonumber \\ 
\hat{h}_{\rm{sp},\vec{r}_i} [\psi] =& v_s  \Big (
(1 + \psi (\vec{r}_i)) c_{\vec{r}_i,A}^{\dagger}c_{\vec{r}_i,A} \nonumber \\ &- (1 + \psi (\vec{r}_i+ \vec{\delta}_1)) c_{\vec{R}_i,B}^{\dagger}c_{\vec{r}_i ,B}\Big) . \nonumber \\ 
\hat{h}_{\rm{ep},\vec{r}_i} [\psi, \phi] =&  \Big (
(1 + \psi (\vec{r}_i)) \phi (\vec{r}_i) c_{\vec{r}_i,A}^{\dagger}c_{\vec{r}_i,A} \nonumber \\ &- (1 + \psi (\vec{r}_i+ \vec{\delta}_1)) \phi (\vec{r}_i  + \vec{\delta}_1) c_{\vec{r}_i,B}^{\dagger}c_{\vec{r}_i,B}\Big).
\end{align}

We now sketch how to obtain the smoothly varying energy current density operator from this lattice model for calculating the energy current density. (i) The local and discrete energy current operator $\hat{J}_{ij}^{E}$ to flow from $\vec{r_j}$ to $\vec{r_i}$ can be identified from the continuity equation 
\begin{align}
    \frac{\partial}{\partial t}\hat{h}_{\vec{r}_i} = \sum_{\vec{r_j}}\hat{J}_{ij}^{E} = \frac{i}{\hbar} \sum_{\vec{r_j}} \left [\hat{h}_{\vec{r}_i}, \hat{h}_{\vec{r}_j} \right], 
\end{align}
as 
\begin{align} \label{eqsup:discreteenergycurrent}
    \hat{J}_{ij}^{E} = \frac{i}{\hbar} \left [\hat{h}_{\vec{r}_i}, \hat{h}_{\vec{r}_j} \right]. 
\end{align}
Note that $\hat{J}_{ij}^{E} = - \hat{J}_{ji}^{E}$ as it should be.
(ii) Being interested in the long-range behavior of the energy current, we find the continuous energy density operator $\hat{h}_{
\psi} (\vec{r})$ and the energy current density $\hat{\vec{J}}{}_{
\psi}^E (\vec{r})$. $\hat{h}_{
\psi} (\vec{r})$ can be obtained from the local Hamiltonian $\hat{h}_{\vec{R}_i}$ as
\begin{align}
    \hat{h}_{
\psi} (\vec{r}) = \sum_{\vec{R
r}_i} \hat{h}_{\vec{r}_i} f (\vec{r} -\vec{r}_i), 
\end{align}
Here, we introduce a smoothing function $f (\vec{r} -\vec{r}_i)$ that decays in length scale $W$, much larger than the lattice constant $a$, but much smaller than $
\sigma$ where $\psi (\vec{r})$ and $
\phi(\vec{r})$ decay. $f(\vec{r})$ is defined to normalize as 
$\int d\vec{r} f(\vec{r})= 1$. 
Similarly $\hat{\vec{J}}{}_{
\psi}^E (\vec{r})$ can be written as
\begin{align} \label{eqsup:contenergycurrentdensity}
    \hat{\vec{J}}{}_{
\psi}^E  (\vec{r}) = \sum_{\vec{r}_i, \vec{r}_j}f \left (\vec{r} - \frac{\vec{r}_i +\vec{r}_j }{2} \right)  \frac{
   \left( \vec{r}_i -\vec{r}_j \right) }{2} \hat{J}_{ij}^{E}. 
\end{align}
The continuity equation $\partial \hat{h}_{
\psi} (\vec{r}) / \partial t + \vec{\nabla} \cdot \hat{\vec{J}}{}_{
\psi}^E (\vec{r}) = 0$ is shown to be valid as the smoothing function changes monotonously in the atomic scale $\sim a$ such that 
the linear-order expansion remains valid,
\begin{align}
\vec{\nabla} f \left(\vec{r} -  \frac{ \vec{r}_i +\vec{r}_j }{2} \right) \approx 
\frac{f(\vec{r} - \vec{r}_j) -f(\vec{r} - \vec{r}_i) }{\vec{r}_i - \vec{r}_j}. 
\end{align}
(iii) Although $\hat{\vec{J}}{}_{
\psi}^E (\vec{r})$ in Eq.~\eqref{eqsup:contenergycurrentdensity} is defined to satisfy the continuity equation, $\hat{\vec{J}}{}_{
\psi}^E (\vec{r})$ has a degree of freedom to add 
$\nabla \times \vec{g} (\vec{r})$ with a continuous function $\vec{g} (\vec{r})$. To uniquely define $\hat{\vec{J}}{}_{
\psi}^E (\vec{r})$, we impose the locality condition~\cite{Cooper1997, Qin2011, Vinkler-Aviv2019}
\begin{align} \label{eqsup:locality}
\hat{\vec{J}}{}_{
\psi}^E (\vec{r})= (1 + \psi (\vec{r}))^2 \hat{\vec{J}}{}_{
\psi = 0}^E (\vec{r}). 
\end{align}
We note from Eqs.~\eqref{eqsup:discreteenergycurrent} and \eqref{eq:supplelatHamiltoniangrav} that $\hat{J}_{ij}^{E}$ contains the terms with $\sim(1 + \psi (\vec{r}_i))(1 + \psi (\vec{r}_j))$ at different locations $\vec{r}_i \neq \vec{r}_j$, which apparently do not fulfill the locality condition Eq.~\eqref{eqsup:locality}. In order to ensure Eq.~\eqref{eqsup:locality} fulfilled, we expand $(1 + \psi (\vec{r}_i))(1 + \psi (\vec{r}_j))$ at $
\vec{r}_{ij} = (\vec{r}_i +\vec{r}_j)/2$ to the linear order as 
\begin{align} \label{eqsup:expansiongravpot}
& (1 + \psi (\vec{r}_i))(1 + \psi (\vec{r}_j)) \nonumber \\ 
&\approx (1+\psi (\vec{r}_{ij}))^2 + \frac{(\vec{r}_i + \vec{r}_j-2\vec{r}_{ij})}{2}  \partial_{\vec{r}_{ij}} 
\left[(1+\psi (\vec{r}_{ij}))^2 \right]. 
\end{align}
Since the smoothing function acts as a delta function on the scale $\sigma$ in Eq.~\eqref{eqsup:contenergycurrentdensity}, the first term in Eq.~\eqref{eqsup:expansiongravpot} results in $(1+ \psi(\vec{r}))^2$, and hence satisfy the locality condition  [Eq.~\eqref{eqsup:locality}].  
This locality condition can be achieved also in the second term of Eq.~\eqref{eqsup:expansiongravpot}
after first replacing $\vec{r}_{ij}$ with $\vec{r}$ and then 
subtracting off the relevant terms in the form $\nabla \times [(1+ \psi (
\vec{r}))^2 \vec{g} (\vec{r}) ] $, allowed by the remaining degree of freedom of $\hat{\vec{J}}{}_{\psi}^E$ from the continuity equation.

We consider this Haldane lattice model with finite size $N_x$ in the $x$ direction, labeled by $i,j=1,\dots,N_x$, but infinitely long in the $y$ direction (cf. Fig.~\ref{Fig:suppleHaldanemodel}). For simplicity, the gravitational (or electrostatic) potential is assumed to vary only in the $x$ direction while it remains constant in the $y$ direction. Periodic boundary conditions are imposed  both in the $x$ and $y$ directions. From Eq.~\eqref{eqsup:contenergycurrentdensity}, the charge  current density $J_{y}^{C} (x)$ at $x$ along the $y$ direction can be written as 
\begin{align} \label{eqsup:latticechargecurrent}
    J_{y}^{C} (x) &= \sum_{i, j=1,\cdots, N_x} \sum_{\alpha \beta} \sum_{n} \int_{- \frac{\pi}{3a}}^{\frac{\pi}{3a}}
    \frac{d k_y}{(2 \pi)} f \left(x - \frac{r_{i \alpha} + r_{j \beta}}{2}\right) \nonumber \\ 
    & \times \phi_{n i \alpha}^{*} (k_y) A^{C}_{i j, \alpha \beta} (k_y)
    \phi_{n j \beta} (k_y). 
\end{align}
Here, $\phi_{n j  \alpha} (k_y)$ is the wave function of band $n$ at position $r_{j \alpha}= \left (\vec{r}_{j, \alpha} \right)_x$  with momentum $k_y$ along the $y$ direction in the presence of the gravitational (electrostatic) potential. 
This wave function can be obtained from the exact diagonalization of the $N_x \times N_x$ lattice Hamiltonian. 
We sum only over occupied states. Moreover, we defined a kernel $A^{C }_{i j,\alpha \beta}\equiv  \sum_{\left (\vec{r}_{i, \alpha}   -\vec{r}_{j, \beta} \right)_y } J^{C}_{i j, \alpha \beta} \left (\vec{r}_{i, \alpha}  -\vec{r}_{j, \beta} \right)_y e^{-i k_y \left (\vec{r}_{i, \alpha}  -\vec{r}_{j, \beta} \right)_y }/2$ for the thermal (charge) response.
The kernel $A^{C}_{i j}$ for the charge response reads
\begin{align}
   A^{C}_{i j}  = &- e \big ( t_1 \delta_{1, y} \left (\sigma_y \cos(k_y \delta_{1, y}) + \sigma_x \sin (k_y \delta_{1, y} ) \right)
   \delta_{i,  j}
    \nonumber \\ & + t_1 \delta_{2, y} (\sigma_y \cos(k_y \delta_{2, y}) + \sigma_x \sin(k_y \delta_{2, y})) 
    \nonumber \\ & \times
    \left (\delta_{i,  j+1} +\delta_{i,  j-1}   \right) \nonumber \\ 
     & - 2 i t_2 \sigma_z \sigma_{2, y} \sin (k_y\sigma_{2, y} ) 
       \left (\delta_{i,  j+1} -\delta_{i,  j-1} \right) \big ). 
\end{align}
The numerical results for the charge current density $J_y^C (x)$ are plotted in Fig.~3(a) and (b). The calculation has been done with the exact diagonalization of the $100 \times 100$ lattice Hamiltonian. 
We use  $\sigma = 10 a$, $\phi_0=0.1 t_1$, $t_1  = t_2=1$,  and the staggered potential $v_s  = (3 \sqrt{3} - 1) t_2$ in the topological phase (left) and $v_s = ( 3 \sqrt{3} +1) t_2$  in the trivial phase (right). The width of the smoothing function is set to $W = 3 a$.
The numerical results match well with the analytic formula Eq.~(11) in the main text.


From the procedure (i), (ii), (iii) stated above,
we also obtain the energy current density along the $y$ direction
\begin{align} \label{eqsup:latticeenergycurrent}
    J_{y}^{E} (x) &= \sum_{i, j=1,\cdots, N_x} \sum_{\alpha \beta} \sum_{n} \int_{- \frac{\pi}{3a}}^{\frac{\pi}{3a}}
    \frac{d k_y}{(2 \pi)} \phi_{n i \alpha}^{*} (k_y) \phi_{n j \beta} (k_y) \nonumber \\ 
    & \times \Big[ f \left(x - \frac{r_{i \alpha} + r_{j \beta}}{2}\right) A^{E (1)}_{i j, \alpha \beta} (k_y)
    \nonumber \\ 
    & + \partial_{x} f \left(x - \frac{r_{i \alpha}+ r_{j \beta}}{2}\right) (r_{i \alpha}- r_{j \beta}) A^{E (2)}_{i j, \alpha \beta} (k_y)\Big ].  
\end{align}
Note that compared with the charge current, Eq.~\eqref{eqsup:latticechargecurrent}, there are additional terms $\sim \partial_{x} f$ arising from the locality condition, Eq.~\eqref{eqsup:locality}. 
The energy kernels $A^{E (1)}_{i j}$, $A^{E (2)}_{i j}$ for the energy response are given by 
\begin{align} 
 A_{i j}^{E (1)} = & -4 t_2^2 \sigma_{3, y} I \sin(2 k_y \sigma_{3, y})  \delta_{i,  j}
 \nonumber \\ & +(2 t_2^2- t_1^2) \sigma_{3, y} I \sin(k_y \sigma_{3, y} ) (\delta_{i, j+1} + \delta_{i, j-1}) 
 \nonumber \\ & - 2 i t_2 v_s \sigma_{3, y} I \sin(k_y \sigma_{3, y} ) (\delta_{i, j+1} - \delta_{i, j-1}) 
 \nonumber \\ & - 2 t_2^2 \sigma_{3, y} I \sin(2k_y \sigma_{3, y} ) (\delta_{i, j+2} + \delta_{i, j-2}) 
 \nonumber \\ & - 2 t_2^2 \sigma_{3, y} I \sin(k_y \sigma_{3, y} ) (\delta_{i, j+3} + \delta_{i, j-3}),
\end{align}
and 
\begin{align} 
 A_{i j}^{E (2)} = & -t_1 t_2 \delta_{1, y} ( \cos( k_y \delta_{1,y} )  \sigma_x - \sin( k_y \delta_{1,y}) \sigma_y ) \delta_{i,  j}
 \nonumber \\& + 2 t_1 t_2 \delta_{1, y} ( \cos(2 k_y  \delta_{1, y} )  \sigma_x +\sin(2 k_y  \delta_{1, y}) \sigma_y)  \delta_{i,  j}
 \nonumber \\
 & + i t_1^2 \sigma_{3, y} \sigma_z \cos(k_y \sigma_{3, y} ) (\delta_{i, j+1} - \delta_{i, j-1}) /4
 \nonumber \\ & - i t_1 v_s \delta_{2, y} (\cos(k_y \delta_{2, y})\sigma_x  - \sin(k_y \delta_{3, y})\sigma_y ) 
 \nonumber \\ & \times 
 (\delta_{i, j+1} - \delta_{i, j-1}) /4
 \nonumber \\ & + t_1 t_2 \delta_{2, y} (\cos(k_y \delta_{2, y})\sigma_x  - \sin(k_y \delta_{2, y})\sigma_y ) 
 \nonumber \\ & \times 
 (\delta_{i, j+1} + \delta_{i, j-1}) /4
 \nonumber \\ &  + t_1 t_2 \delta_{1, y} (\cos(k_y \delta_{1, y})\sigma_x  - \sin(k_y \delta_{1, y})\sigma_y ) 
 \nonumber \\ & \times 
 (\delta_{i, j+2} + \delta_{i, j-2})/4
 \nonumber \\ &  + t_1 t_2 \delta_{2, y} (\cos(k_y \delta_{2, y})\sigma_x  - \sin(k_y \delta_{2, y})\sigma_y ) 
 \nonumber \\ & \times 
 (\delta_{i, j+3} + \delta_{i, j-3})/12,
\end{align}
respectively. 

The numerical results for the energy current density $J_y^E (x)$ are plotted in Fig.~3(c) and (d) in the main text. 
We used parameters $N_x = 100$, $\sigma = 10 a$, $\psi_0=0.1$, $t_1  = t_2=1$, and the staggered potential $v_s  = (3 \sqrt{3} - 1) t_2$ in the topological phase (left) and $v_s = ( 3 \sqrt{3} +1) t_2$  in the trivial phase (right).
The width of the smoothing function is set to $W = 3 a$. The numerical results for $J_y^E (x)$ can be nicely fitted to $J_y^E = C \frac{v^2}{48 \pi} \partial_x^3 \psi (x)$, but the coefficient $C$ is highly non-universal depending on the lattice parameters, in contrast with the electrical response. 

\section{Response to temperature or chemical potential bump}
In this section, we analytically calculate the Hall response to a smoothly varying chemical potential, $\mu(\vec{r})$, and temperature, $T(\vec{r})$, in the absence of $\psi (\vec{r})$ and $\phi (\vec{r})$ employing the gradient expansion approach.
To be able to change locally the temperature and the chemical potential of the system, we couple weakly to each lattice site $i$ of the honeycomb model, Eq.~(8) in the main text, a wire with a chemical potential $\mu_i$ and temperature $T_i$, see Fig.~2 in the main text, using a tunneling contact of strength $V$. For simplicity, the chemical potential and the temperature are assumed to vary in the $x$ direction while being constant in the $y$ direction. In this section, we neglect phonon baths attached to the system, but the effect of phonons will be discussed in Sec.~\ref{supplesec:phonon}.

In the continuum limit one can describe the coupling to these wires  by the Hamiltonian
\begin{align} \label{wirecoupling}
\hat{H}_t=\int d^2\vec r\, \sum_{q,\sigma}   \left( \epsilon_q d^\dagger_{\sigma,\vec r,q} 
d_{\sigma,\vec r,q} + V d^\dagger_{\sigma,\vec r,q}\Psi_\sigma(\vec r)  + h.c.\right),
\end{align}
where $d^\dagger_{\sigma,\vec r,q}$ creates an electron with energy $\epsilon_q$ in a wire attached to the point $\vec r$ and sublattice $\sigma$ of the Haldane model.
Chemical potentials $\mu_i$ and temperatures $T_i$ of the wires vary in the real space: $\mu_i = \mu (\vec{r}_i)$ and $T_i = T(\vec{r}_i)$. This information is, however, not encoded in the Hamiltonian but in the Fermi function $f(\omega, \vec r)$ describing incoming electrons of the wire attached at position $\vec r$. We use the continuum limit of the Haldane model, Eq.~(9).

The process of tunneling back and forth between the system and the wires is encoded in the retarded and advanced part of the self energy
\begin{align} \label{appen:selfenergyretwire}
\Sigma_{\sigma \vec{k}, \sigma' \vec{k'}}^{R/A} &= \delta_{\sigma \sigma'}\sum_{q} \sum_\vec{r} e^{-i \vec{k}\cdot \vec{r}}  \frac{|V|^2 }{\omega - \epsilon_q \pm i \eta} e^{i \vec{k'}\cdot \vec{r}} \nonumber \\
&= \delta_{\sigma \sigma'} \delta_{\vec{k} \vec{k'}}\sum_{q} \frac{|V|^2}{\omega - \epsilon_q \pm i\eta}\nonumber \\
&\approx \mp i \delta_{\sigma \sigma'}\delta_{\vec{k} \vec{k'}} \Gamma
\end{align}
where we assumed a constant density of states, $N_F$, of the wires with a large bandwidth.
The tunneling rate $\Gamma \equiv \pi N_F |V|^2$ is thus frequency independent. 
Note that the retarded and advanced part of the self energy are translationally invariant and momentum independent. The lesser part of the self energy, on the other hand, does depend on position and therefore also on momentum
\begin{align}\label{appen:selfenergylesserwire}
\Sigma_{\sigma \vec{k}, \sigma \vec{k'}}^{<}&=  \sum_{q} \sum_\vec{r}  e^{-i \vec{k}\cdot \vec{r}} |V|^2
\left (2\pi i \delta (\omega - \epsilon_q) \right )f (\epsilon_q, \vec{r})
 e^{i \vec{k'}\cdot \vec{r}}
\nonumber \\ 
&= 2 i \Gamma  \sum_\vec{r} f (\omega, \vec{r}) e^{-i (\vec{k} - \vec{k'}) \cdot \vec{r}}, 
\end{align}
reflecting the momentum transfer to the wires. Employing the Keldysh equations, 
\begin{align} 
G^{R/A} (\omega ) &= g^{R/A} (\omega) + g^{R/A} (\omega) \Sigma^{R/A} (\omega) G^{R/A} (\omega),  \\ 
G^{<} (\omega) &= G^{R} (\omega) \Sigma^{<} (\omega) G^{A} (\omega),
\end{align} 
the full dressed retarded and advanced Green's function, and the lesser Green's function are given by
\begin{align} \label{supeq:GreenFunRetarded}
G^{R/A}_{\sigma' \vec{k'}, \sigma \vec{k}} (\omega) 
&= \left (\frac{1}{\omega - h \left [ \vec{k} \right ] \pm i \Gamma} 
 \right )_{\sigma' \sigma} \delta_{\vec{k} \vec{k'}}, \\ 
G^{<}_{\sigma' \vec{k'}, \sigma \vec{k}} (\omega) &= 2 i \Gamma \sum_{\vec{r}} f(\omega, \vec{r}) e^{i (\vec{k} - \vec{k'})\cdot \vec{r}} \nonumber \\ \times &
\left ( \frac{1}{\omega - h\left [ \vec{k'} \right ] + i \Gamma} 
 \frac{1}{\omega -  h\left [ \vec{k} \right ]  - i \Gamma}
\right )_{\sigma' \sigma}. \label{supeq:GreenFunlesser}
\end{align}

\label{appen:statisticalresponse}

\subsection{Response to chemical potential bump}

We next consider the charge Hall current in response to a chemical potential bump with $\mu (x) = \mu_0 e^{-x^2/\sigma^2}$. The information of $\mu (x)$ is encoded in the Fermi function $f (\omega, \vec{r})=f(\omega-\mu(x))$ in the lesser Green's function of Eq.~\eqref{supeq:GreenFunlesser}. 

Plugging Eqs.~\eqref{supeq:GreenFunRetarded} and \eqref{supeq:GreenFunlesser} into Eq.~\eqref{currentexpvalue}, we obtain the electric current density $J^{C}_{y, \mu} (\vec{r})$ flowing along the $y$ direction 
\begin{align} \label{supeq:elecurchebump}
J_{y, \mu}^{C} (\vec{r}) = &-2 \Gamma e \int \frac{d \omega}{2\pi}  \int \frac{d\vec{K}}{(2\pi)^2} \int \frac{d\vec{q}}{(2\pi)^2} 
  \nonumber \\ & \times \int d\vec{r_1} f (\omega, \vec{r}_1)  e^{i \vec{q} \cdot (\vec{r} - \vec{r}_1)} 
\textrm{Tr}\Big [ \left ( v  \sigma_y
+ 2 \lambda^2 \sigma_z K_y\right ) 
\nonumber \\ & \times 
\frac{1}{\omega - h (\vec{K}+\vec{q}/2) + i \Gamma }
 \frac{1}{\omega - h (\vec{K} - \vec{q}/2)  - i \Gamma }
 \Big ].
\end{align}
Being interested in smooth potentials, we expand the term inside the trace in $q_x$.
The leading contribution comes from the linear order in $q_x$, and the direct calculation of the integral of the linear term in $q_x$ results in 
\begin{align} \label{eqsup:statisticalforce_current}
J_{y, \mu}^{C} (\vec{r}) &=  \frac{\hbar \Gamma e v^2 \partial_x \mu}{ \pi^2} \int \frac{dK K (M- \lambda^2 K^2)}{
\left ( \left ( \lambda^2 K^2 + M\right)^2 + v^2 K^2 + \Gamma^2 \right)^2}.
\end{align}
We have also expanded $\partial_x f (\omega, \vec{r}) \approx \frac{\partial f(\omega)}{\partial \mu}|_{\mu=0} \partial_x \mu(x)$ around zero chemical potential.
As $G^{R/A}$ in Eq.~\eqref{supeq:elecurchebump} is always finite for $|\omega| \ll M$ and $\Gamma \to 0$, it follows directly from Eq.~\eqref{supeq:GreenFunlesser} that 
$G^{<} \propto \Gamma$ and thus $J_{y, \mu}^{C} \propto \Gamma$ for $\Gamma \to 0$.
The physical interpretation of this result is that some charge tunnels between the attached wires through the gapped topological insulator. This type of transport does, however, vanish in the limit $\Gamma \to 0$. We conclude that spatially varying chemical potentials do  not induce any topological currents in Chern insulators, implying that the Einstein relation is not valid. 

\begin{figure} 
\includegraphics[width=0.7\columnwidth]{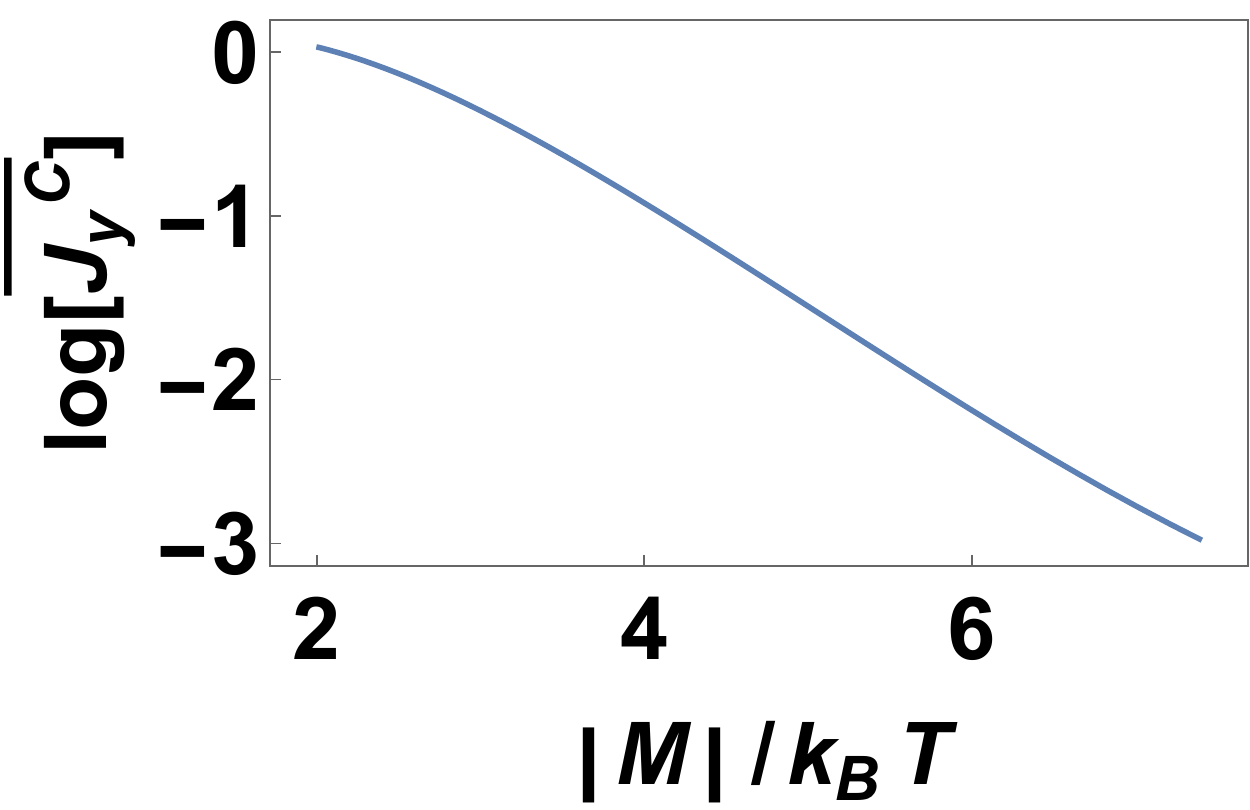} 
\caption{{\bf Exponential suppression of the electric current $J_{y, \mu}^C$ by energy gap.} The logarithm of the normalized electric current $\overline{J_{y, \mu}^C} \equiv J_{y, \mu}^C [k_B T ] / J_{y, \mu}^C [k_B T = |M|=1]$ is plotted as a function of  $|M|/ (k_B T)$. We choose $\Gamma/ |M| = 0.005$. In the intermediate scale of $1< |M|/(k_B T) < \log (|M|/\Gamma)$, $J_{y, \mu}^C$ is exponentially suppressed by energy gap. 
}
\label{Fig:exponentialsuppression}
\end{figure}

We next consider the effect of the finite temperature. 
The logarithm of the function $J_{y, \mu}^C \left[\frac{k_B T}{|M|} \right] / J_{y, \mu}^C \left [ \frac{k_B T}{|M|} = 1 \right]$ is plotted as a function of $|M|/ (k_B T)$ in Fig.~\ref{Fig:exponentialsuppression}. We choose $\Gamma/ |M| = 0.005$. 
In the intermediate scale of temperatures $|M| / \log (|M|/\Gamma)< k_B T< |M| $ (equivalently, $|M| e^{-|M|/(k_B T)}<\Gamma $ and $k_B T < |M|$), the electric current exponentially is suppressed as $J_{y, \mu}^C \approx \exp (- \alpha |M| / k_B T)$ with numerical constant $\alpha \simeq 0.63$. This exponentially small contribution originates from excitations above the energy gap.





\begin{widetext}
\subsection{Response to temperature bump}
Finally, we consider the Hall energy current in response to a temperature bump, $T (x)$, in the absence of the gravitational potential.

Plugging Eqs.~\eqref{supeq:GreenFunRetarded} and \eqref{supeq:GreenFunlesser} into Eq.~\eqref{supeq:energycurrentexp} (with $\psi (\vec{r})=0$), we obtain the energy Hall current $J_{y, T}^{E} (\vec{r})$
\begin{align}
J_{y, T}^{E} (\vec{r}) 
&=  2 \Gamma \int \frac{d^2 \vec{K}}{(2\pi)^2} \int \frac{d^2 \vec{q}}{(2\pi)^2} \int \frac{d \omega}{2\pi} 
\int d\vec{r}_1 f(\omega, \vec{r}_1) e^{i \vec{q} \cdot (\vec{r} - \vec{r}_1)} \textrm{Tr} \bigg [ \frac{1}{\omega - h (\vec{K}+\vec{q}/2) + i \Gamma }
 \frac{1}{\omega - h (\vec{K} - \vec{q}/2) - i \Gamma }  \nonumber \\ &\times 
\bigg \{  \left ( v^{2} + 2 \lambda^2 \left (M + \lambda^2 \vec{K}^2   \right)
\right ) K_y    - \frac{i}{4} \left (v^2 \sigma_z - 2 v \lambda^2 \left( \sigma_x K_x + \sigma_y K_y \right) \right) q_x \bigg \}  
 \bigg ]. 
\end{align}
The leading contribution comes from the term linear in $q_x$ and results in 
\begin{align} \label{subeq:leadingcontribution}
J_{y, T}^{E} (\vec{r}) &= 2 \Gamma \int \frac{d K }{2\pi} \int \frac{d \omega}{2\pi} \int \frac{d^2 \vec{q}}{(2\pi)^2} 
\int d\vec{r'} 
\frac{i K (\lambda^2 K^2 - M) v^2 q_x e^{i \vec{q}\cdot  \left (\vec{r} - \vec{r'} \right)} f(\omega, \vec{r'})  \omega}{\left[ \left (\lambda^2 K^2 +M\right)^2 + K^2 v^2 + \Gamma^2 \right]^2 - 2\omega^2 \left [ \left (\lambda^2 K^2 +M\right)^2 + K^2 v^2 - \Gamma^2 \right] + \omega^4} \nonumber \\ 
& = 2 \Gamma   \int \frac{d K }{2\pi} \int \frac{d \omega}{2\pi} \partial_x f(\omega, x)
\frac{K (\lambda^2 K^2 - M) v^2  \omega}{\left[ \left (\lambda^2 K^2 +M\right)^2 + K^2 v^2 + \Gamma^2 \right]^2 - 2\omega^2 \left [ \left (\lambda^2 K^2 +M\right)^2 + K^2 v^2 - \Gamma^2 \right] + \omega^4}.
\end{align}
 The chain rule $\partial_x f(\omega, x) = \partial_x T \partial_T f (\omega)$ and the Sommerfeld expansion yield
\begin{align} \label{supeq:sommerfeld}
J_{y, T}^{E} (\vec{r}) &=    \partial_x T (x)  \frac{\partial}{\partial_T} \left ( \int \frac{d \omega}{2\pi} \int \frac{d K }{2\pi}
\frac{ 2 \Gamma  K (\lambda^2 K^2 - M) v^2  \omega f(\omega)}{\left[ \left (\lambda^2 K^2 +M\right)^2 + K^2 v^2 + \Gamma^2 \right]^2 - 2\omega^2 \left [ \left (\lambda^2 K^2 +M\right)^2 + K^2 v^2 - \Gamma^2 \right] + \omega^4} \right) \nonumber \\ 
& =  \partial_x T (x)    \frac{\partial}{\partial_T} \left (\frac{\pi^2 (k_{B} T)^2 }{6h} \left ( \int \frac{d K }{2\pi}
\frac{2 \Gamma  K (\lambda^2 K^2 - M) v^2}{\left [\left (\lambda^2 K^2 +M\right)^2 + K^2 v^2 + \Gamma^2 \right]^2 } + O \left (\frac{ (k_B T)^2}{|M|^2}\right)  \right) \right)  \nonumber \\ 
& \approx \frac{\pi^2 k_B^2 T}{3h} \partial_x T \left (\int \frac{d K }{\pi} 
\frac{ \Gamma  K (\lambda^2 K^2 - M) v^2}{\left [\left (\lambda^2 K^2 +M\right)^2 + K^2 v^2 + \Gamma^2 \right]^2 } \right ).
\end{align}
\end{widetext}
In the last expression, we neglect the higher order terms $O ((k_B T / |M|)^2 )$, which in turn yields the exponential suppression by the energy gap. 
The direct integration over momentum $K$ in the limit $\Gamma \to 0$, and $|M| \ll v^2/\lambda^2$ leads to 
\begin{align}
  J_{y, T}^{E} (\vec{r})  \approx -\frac{\pi^2 k_B^2 T}{6h} \frac{\Gamma}{\pi M} \left (\partial_x T + \frac{v^2}{6 M^2} \partial_x^3 T + O(\partial^5_x T)\right).
\end{align}
Here, the gradient term $\propto \partial_x T$ is directly from Eq.~\eqref{supeq:sommerfeld} while the third derivative term $\propto \partial^3_x T$ is the next leading contribution obtained from the cubic term $q_x^3$ in the gradient expansion using the similar procedure as in Eqs.~\eqref{subeq:leadingcontribution} and \eqref{supeq:sommerfeld}. Our calculation confirms the absence of a thermal Hall response in the weak coupling limit, $\Gamma \to 0$. Note that also all contributions proportional to higher derivatives of $T(x)$ vanish in this limit.

\section{Temperature imprinted by the attached wires} \label{sec:temperatureimprint}

\begin{figure} 
\centering \captionsetup[subfloat]{labelfont=bf}
\subfloat[]{\includegraphics[width=0.8\columnwidth]{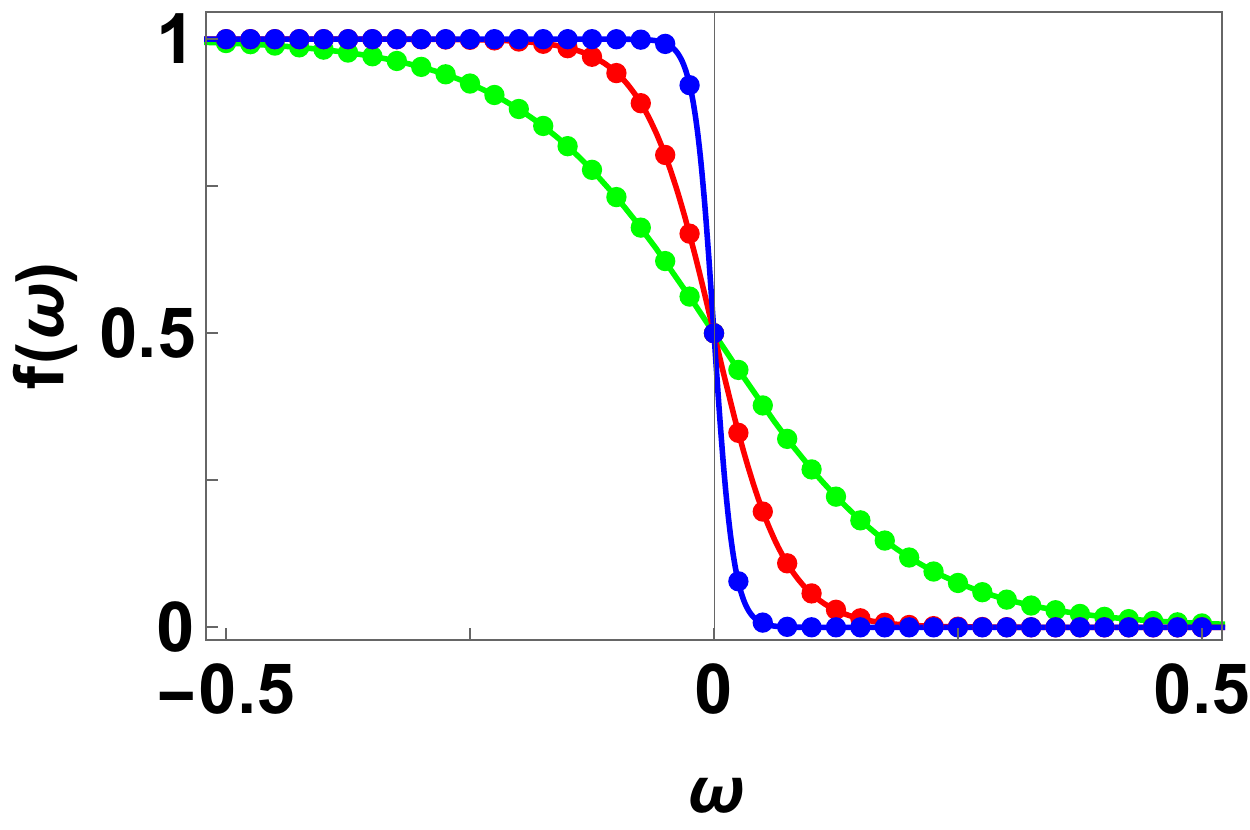}}
\hspace{0.07\columnwidth}
\subfloat[]{\includegraphics[width=0.8\columnwidth]{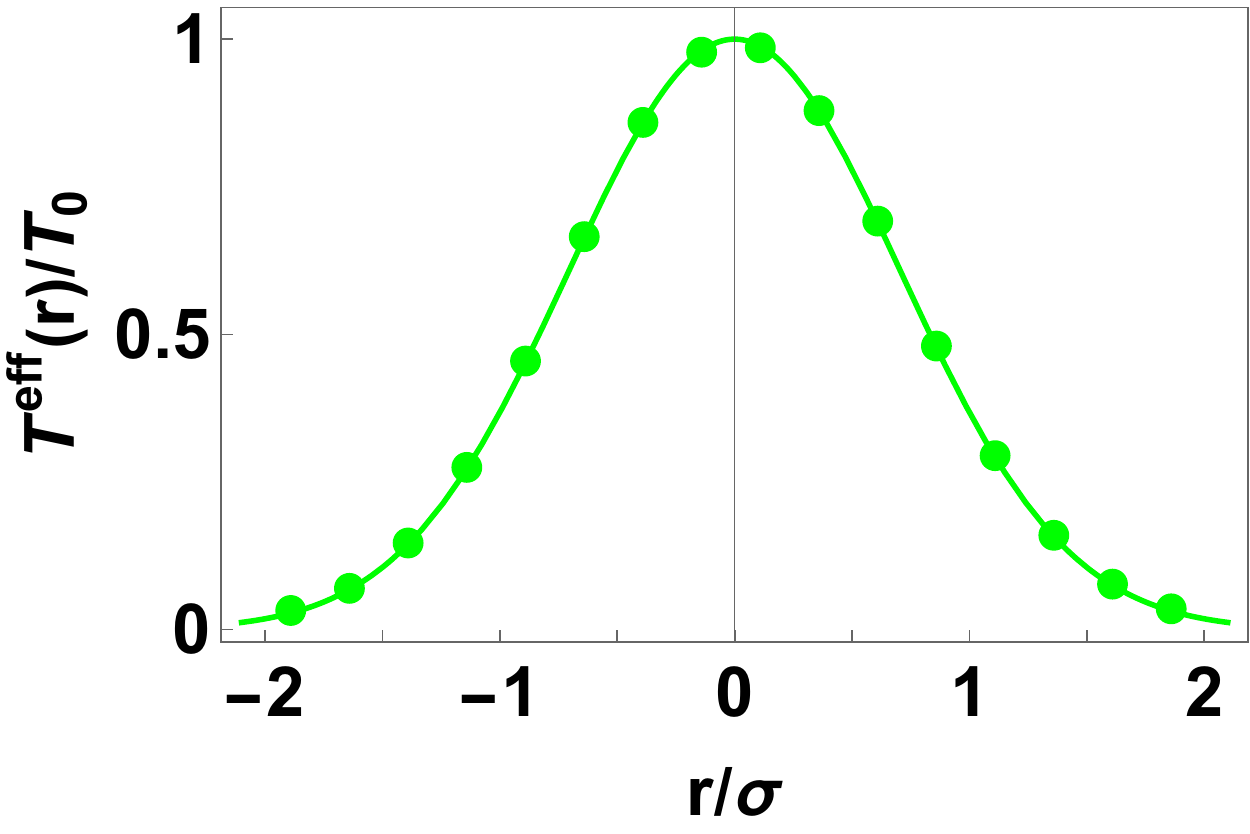}}
\hspace{0.07\columnwidth}
\subfloat[]{\includegraphics[width=0.8\columnwidth]{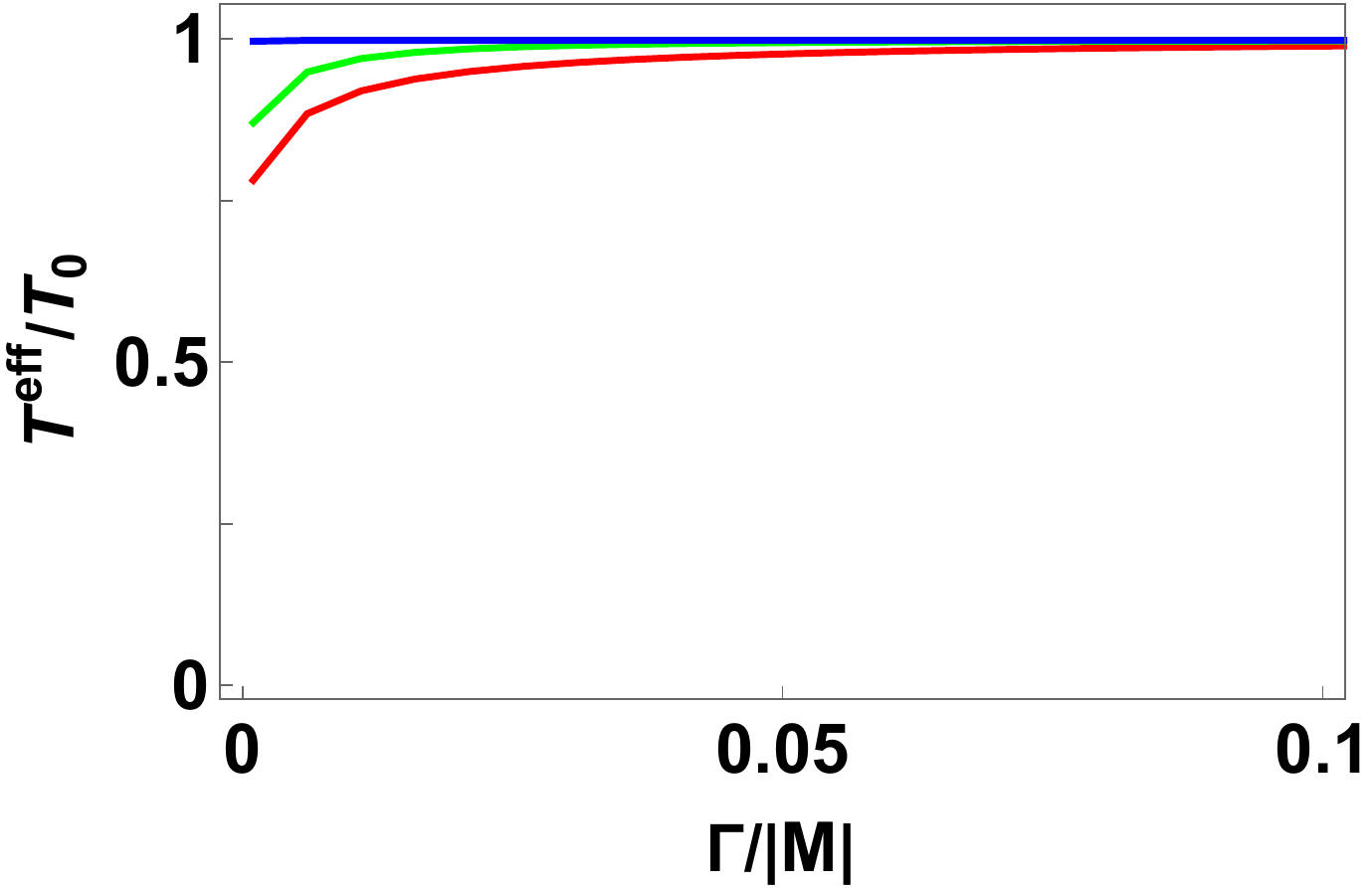} }
\caption{{\bf Temperature imprint onto a Chern insulator by a temperature profile $T(x) = T_0 e^{-x^2/\sigma^2}$ of attached wires (Fig.~2 in the main text).} 
(a) Local distribution function $f_{x}^{\text{eff}} (\omega)$ (filled dots) of the Chern insulator and the distribution $f (\omega, x)$ (solid curves) of the attached wires
 with three different positions $x$; $x/\sigma = 0$ (Green), $x/\sigma= 1$ (red), and $x/\sigma= 1.5$ (blue). Remarkably, two distribution functions are in excellent agreement, showing that the local temperature of the Chern insulator is indeed imprinted by the temperature profile of the attached wires. (b) The local temperature $T^{\text{eff}} (x) = T_{\text{th}} (x)$ (green dots) of the Chern insulator at $\vec{r}$ and 
the temperature profile $T(x) $ (the green curve) of the wires as a function of $x$. 
(c) $T^{\rm{eff}} (x =0) = T_{\rm{th}} (x =0)$ as a function of $\Gamma$ with three different $T_0$; $T_0 = 0.05$ (blue), $T_0 = 0.1$ (green), and $T_0 = 0.2$ (red). $T^{\rm{eff}}$ starts to deviate from $T_0$ in $\Gamma \sim |M| \exp (-|M|/ (k_B T_0))$ due to thermal excitations across the gap $|M|$. Parameters: $v = \lambda = - M = 1$, $\sigma = 32 a$, and $N = 129$ for (a-c), and $\Gamma = 0.1 |M|$ for (a-b).  }
\label{Fig:Localtemp}
\end{figure}

In this section we show numerically that the local temperature of the Chern insulator is determined
with high precision by the temperature of the locally attached wires under the conditions specified below.
To show this, we consider the local distribution function of electrons in the Chern insulator defined by
\begin{align} \label{effectivedist}
    f_{\vec{r}}^{\text{eff}} (\omega) = - \frac{G^<_{\vec{r},\vec{r}} (\omega)}{(G^R - G^A) (\omega)}. 
\end{align}
The retarded and advanced Green function of the system are simply given by
$G^{R/A}_{\sigma' \vec{k'}, \sigma \vec{k}} (\omega) 
= \left (\frac{1}{\omega - h \left [ \vec{k} \right ] \pm i \Gamma} 
 \right )_{\sigma' \sigma} \delta_{\vec{k} \vec{k'}}$, where the only effect of the coupling to the attached wires is the broading induced by the term $i \Gamma$. The lesser Green function, $G^{<}$, in contrast, is not translationally invariant but can be computed exactly from $f(\omega,\vec r)$ and $G^{R,A}$ in position space
\begin{align} \label{supplelesserGreenFun}
G^{<}_{\vec r_j, \vec r_i}(\omega) =& 2 i \Gamma  \sum_{\vec r_n} G^R_{\vec r_j,\vec r_n}(\omega) f(\omega, \vec{r}_{n}) G^A_{\vec r_n,\vec r_i}(\omega).
\end{align} Furthermore, we study the value of the local temperature defined by the following procedure:
we attach to position $\vec{r}$ of the system a quantum wire (the `thermometer') by a tunneling contact with temperature $T_{\rm{th}}$. The thermometer is in local thermal equilibrium with the system if the energy current  into or from the thermometer vanishes, $J^E_{\rm{th}} (\vec{r})$. Here the energy current is obtained from 
the Meir-Wingreen formula~\cite{Meir1992} as
\begin{align} \label{currenttoThermometry}
    J^E_{\rm{th}} (\vec{r}) &= \frac{ie}{h} \int d\omega \,\omega \,\rho_{\rm{th}} |t_{\rm{th}}|^2
 \nonumber \\ 
 &
    \times
    \textrm{Tr} \left [f_{\rm{th}} (\omega) (G^R - G^A) (\omega) + G^<_{\vec{r},\vec{r}} (\omega) \right] 
    \nonumber \\
   &  = \frac{ie}{h} \int d\omega \omega \rho_{\rm{th}} |t_{\rm{th}}|^2 \nonumber \\ &
   \times \textrm{Tr} \left [ (f_{\rm{th}} (\omega) - f_\vec{r}^{\text{eff}} (\omega))(G^R - G^A) (\omega)\right],
\end{align}
where $f_{\rm{th}} (\omega)=1/ [\exp \left (\omega /  T_{\rm{th}}\right) + 1]$ is the distribution function of the thermometer with constant density of states $\rho_{\rm{th}}$ and tunneling rate $t_{\rm{th}}$.
Therefore, the effective local temperature $T^{\rm eff}(\vec r)=T_{\rm th}$ is determined from the condition 
$J^E_{\rm{th}} (\vec{r})=0$.
This thermometer should not be confused with the attached wires considered in the paper so far. 

The distribution function $f_{\vec{r}}^{\text{eff}} (\omega) $ of the Chern insulator can be calculated by discretizing the system as a $N \times N$ square lattice and computing the Green's functions $G^<_{\vec{r},\vec{r}}$, $G^R$, and $G^A$, explicitly, see Eq.~\eqref{supplelesserGreenFun}.
The dots in Fig.~\ref{Fig:Localtemp}(a), show the effective distribution function $ f_{\vec{r}}^{\text{eff}} (\omega)$ at three different positions $\vec r$ obtained from Eq.~\eqref{effectivedist} assuming that the temperature profile of the attached wires is given by $T (x) = T_0 e^{-r^2/\sigma^2}$. The solid lines is simply given by the Fermi function $f (\omega, \vec{r})=1/(\exp(\omega/T(\vec r))+1)$. The perfect agreement shows that the temperature of the Chern insulator matches indeed the local temperature of the wires in this example for low $T$ and $\sigma$ much larger than the correlation length $\xi=v/|M|$ of the Chern insulator.  Fig.~\ref{Fig:Localtemp}(b) shows the comparison of the temperature profile $T(\vec r)$ of the attached wires (solid line) to the effective temperature $T^{\rm eff}(\vec r)$ obtained from the attached thermometer described above.

The fact that the temperature of the wire is directly imprinted onto our Chern insulator can be understood by inspecting the formula for $G^<$,  Eq.~\eqref{supplelesserGreenFun}.
Provided that (i) all temperatures are much smaller than the gap $T(\vec r) \ll |M|$, the main contribution of the integral in Eq.~\eqref{currenttoThermometry} arises within the frequency range of $\omega \in (-T_0, T_0)$ inside the gap in which the spectral function $A (\omega) = - \textrm{Im} \left [(G^R - G^A) (\omega) \right] / \pi$ is proportional to $\Gamma$. 
In this frequency range the retarded and advanced Green functions decay on the length scale  $\xi =v/|M|$. Thus  $G^<_{\vec r,\vec r}$ is only sensitive to distribution functions in the proximity of $\vec r$. As we considered a case where 
 $\xi $  is much smaller than a length scale $\sigma$ over which the temperature profile varies, the local $G^<$  is
 thus only affected by temperatures very close to $T(\vec r)$. This explains the perfect agreement of the measured temperature to the 
 temperature of the attached wires.
 
These arguments remains valid as long as thermal excitations across the gap -- which are not localized on the length scale $\xi$ -- can be neglected. They are exponentially suppressed with $e^{-|M|/T_0}$. 
But, when $\Gamma$ is extremely small such that 
$\Gamma/|M| \lesssim e^{-|M|/T_0} $, the dominant contribution to the integral in Eq.~\eqref{currenttoThermometry}
does not arise from inside the gap, but rather from thermal excitations above the gap. These thermal excitations in turn renders the local temperatures of the Chern insulators to deviate from the temperature profiles of the wires. 
In Fig.~\ref{Fig:Localtemp}(c) we show the ratio of the measured temperature $T^{\rm eff}$ at $\vec r=0$ and the local temperature of the wire as a function of $\Gamma$ for three different temperatures. Small deviations between the two temperatures for small $\Gamma$ are exponentially suppressed when the temperature is lowered.

 In conclusion, we have shown that under the conditions specified above, the temperature profile of the wires is successfully imprinted onto the Chern insulator.
 Therefore, a genuine response to a temperature or chemical potential profile can be obtained from the model with attached wires.

\section{Violation of the Luttinger relation in a phonon-coupled Chern insulator}
\label{supplesec:phonon}
 In this section we consider a model where additionally we attach phonon baths to a Chern insulator 
and demonstrate that the Luttinger relation is invalid. 
Attaching the phonon baths has two advantages. First, the model leads to equilibration of the system even in the absence of any attached wires. Second, the phonon bath allows to imprint a temperature profile onto the system in a way which is much closer to an experimental setting, where one would use, e.g., a laser to heat the system locally. If such an experiment is done on an insulator, phonons provide the dominant equilibration channel.

To describe a Chern insulator with a simpler model, we consider a discretized version of the continuum model
defined in Eq.~\eqref{systemHamiltonian} realized on the square lattice (avoiding the longer-ranged hoppings of the Haldane model). 
The model contains two orbitals $\sigma = \uparrow, \downarrow$ per lattice site $\vec r_i$. It includes two nearest hopping terms and an on-site term
\begin{align} \label{squarelatticeHamiltonian}
H &= H_1 + H_2 + H_3, \nonumber \\ 
H_1 &= \frac{i \hbar v}{2a} \sum_{\vec{r}_i} \left ( \Psi^{\dagger}_{\vec{r}_i + a \hat{x}} \sigma_x \Psi_{\vec{r}_i} + 
\Psi^{\dagger}_{\vec{r}_i + a \hat{y}} \sigma_y \Psi_{\vec{r}_i} - \textrm{H.c.} \right) ,\nonumber \\ 
H_2 &= -\frac{\lambda^2}{a^2} \sum_{\vec{r}_i} \left ( \Psi^{\dagger}_{\vec{r}_i + a \hat{x}} \sigma_z \Psi_{\vec{r}_i} + 
\Psi^{\dagger}_{\vec{r}_i + a \hat{y}} \sigma_z \Psi_{\vec{r}_i} + \textrm{H.c.} \right), \nonumber \\ 
H_3 &= \left ( M  + 4 \frac{\lambda^2}{a^2} \right) \sum_{\vec{r}_i}\Psi_{\vec{r}_i}^{\dagger}\sigma_z \Psi_{\vec{r}_i}, 
\end{align} 
Here $a$ is the lattice constant. 
The two-component spinor $\Psi_{\vec{r}_i }^{\dagger} = (\Psi_{\vec{r}_i  \uparrow}^{\dagger}, \Psi_{\vec{r}_i \downarrow}^{\dagger})$ creates electrons  in the two orbitals, $\uparrow$ and $\downarrow$, at site $\vec{r}_i$. 
Note that the Fourier transform of 
Eq.~\eqref{squarelatticeHamiltonian} is identical with Eq.~\eqref{systemHamiltonian}
in the continuum limit, $\vec{k} \rightarrow 0$. 
At $-4 \lambda^2 / a^2 < M  < 0$ ($-8 \lambda^2 / a^2 < M  < -4 \lambda^2 / a^2 $), the system is in a topological phase with Chern number $c_R- c_L =-1$ ($+1$) while at $M>0$ or $M<-8 \lambda^2 / a^2$, the system is in a trivial phase. At the quantum phase transition to the trivial phase at $-4 \lambda^2 / a^2 = M$, the gap closes at $\vec{k} = (\pm \pi/a, 0)$ and $\vec{k} = (0, \pm \pi/a)$ while it remains finite at $\vec{k} = (0,0)$ point. 

\begin{figure} 
\centering \captionsetup[subfloat]{labelfont=bf}
\subfloat[]{\includegraphics[width=0.8\columnwidth]{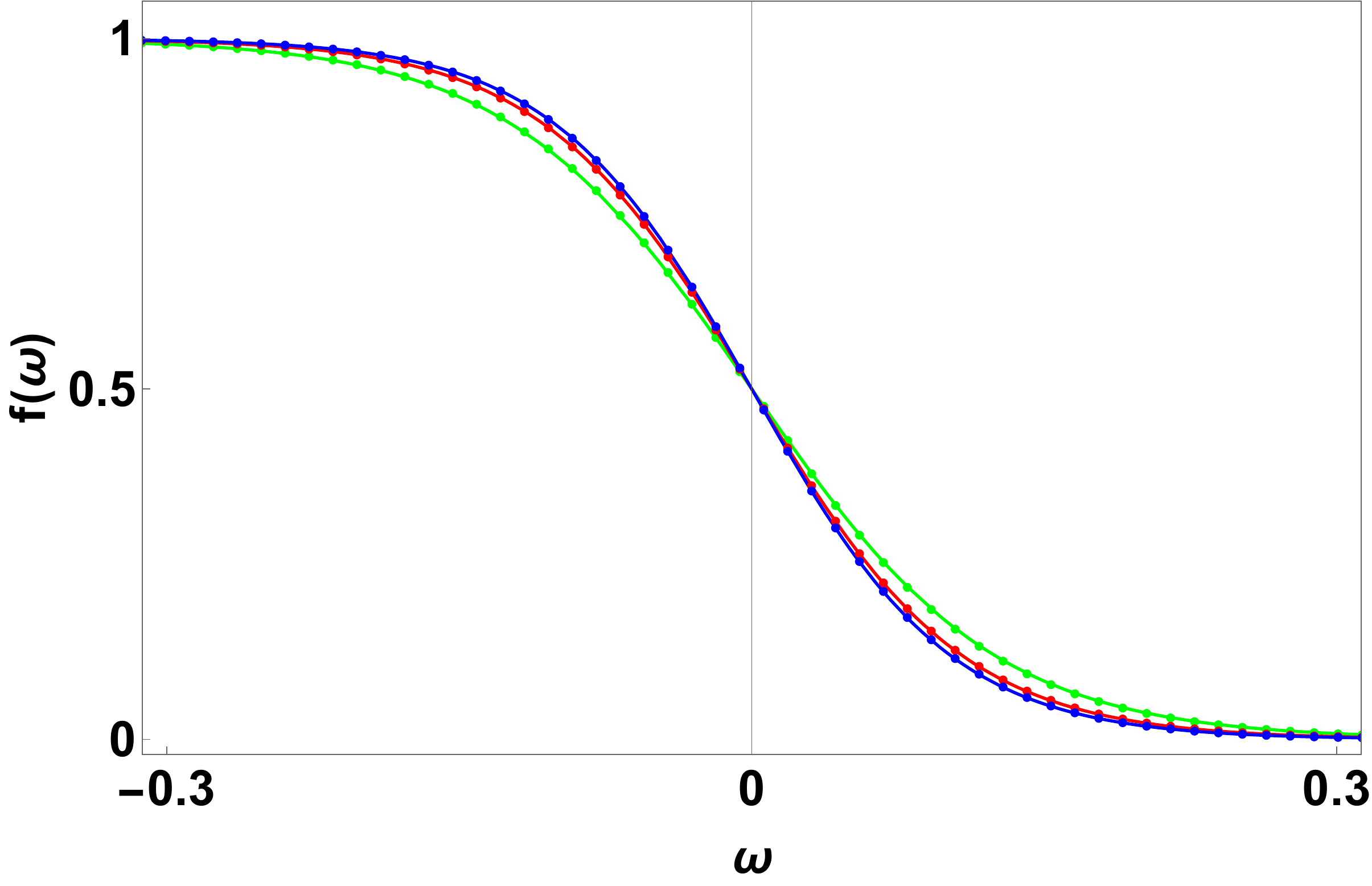}}
\hspace{0.07\columnwidth}
\subfloat[]{\includegraphics[width=0.8\columnwidth]{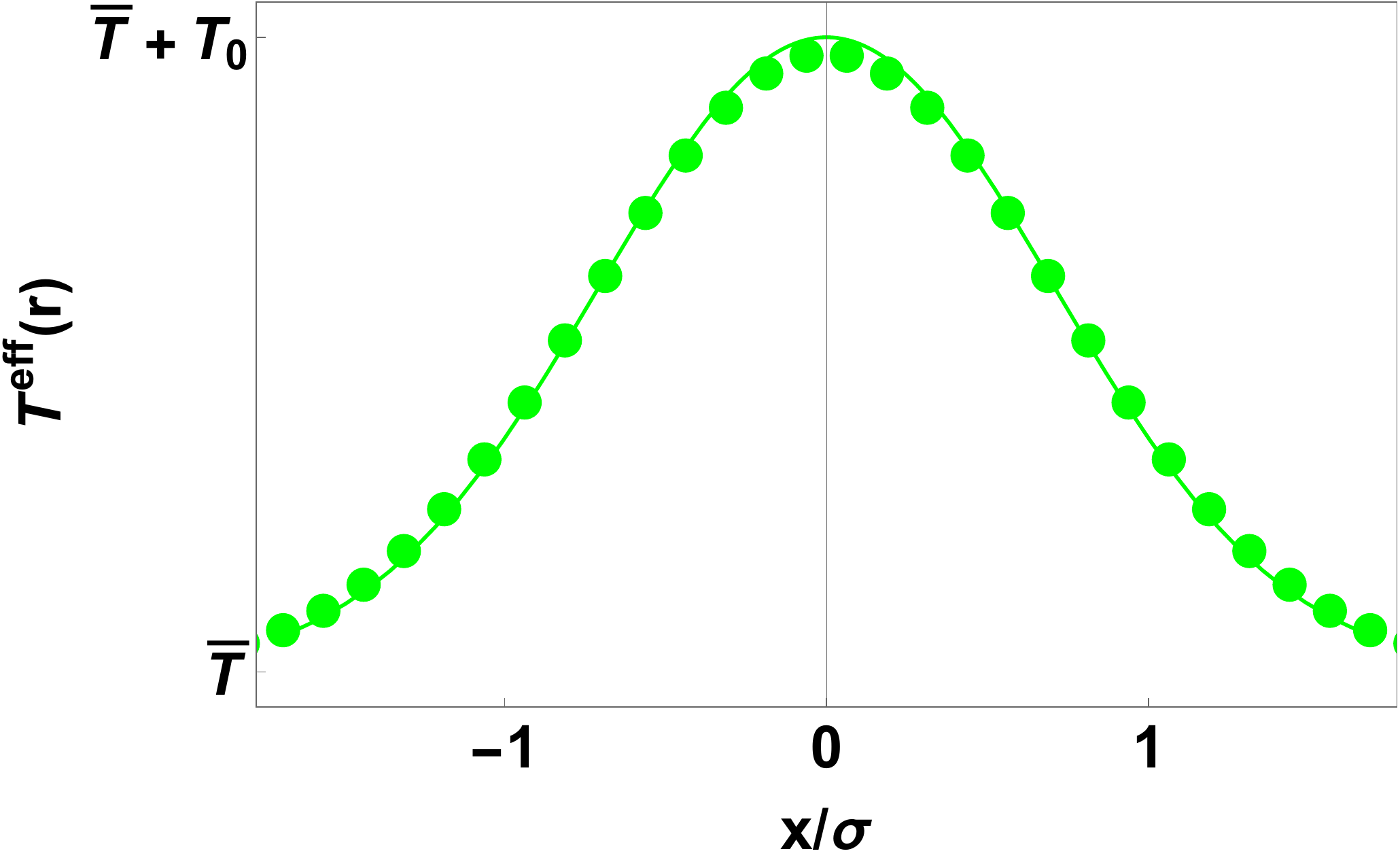}}
\caption{{\bf Local effective temperatures of a Chern insulator coupled to phonon baths with a temperature profile $T(x) = \overline{T} + T_0 e^{-x^2/\sigma^2}$.} 
(a) Local distribution function $f_{x}^{\text{eff}} (\omega)$, defined in Eq.~\eqref{effectivedist}, (filled dots) of the Chern insulator and the distribution $f (\omega, x)$ (solid curves) of the phonon baths with three different positions; $x/\sigma = 0$ (Green), $x/\sigma= 1$ (red), and $x/\sigma= 1.5$ (blue). $\overline{T}/|M| = 0.05$ and $T_0/ |M| = 0.0125$ are chosen. Remarkably, two distribution functions are in perfect agreement, showing that the local temperature of the Chern insulator is indeed imprinted by the temperature profile of the phonon baths. (b) The local temperature $T^{\text{eff}} (x)$ (the green dots) of the Chern insulator at $x$, determined such that $ J^E_{\rm{th}} (x)$ vanishes in Eq.~\eqref{currenttoThermometry}, and 
the temperature profile $T(x) $ (the green curve) of the phonons as a function of $x$. 
Parameters: $v = \lambda = - M = 1$, $\sigma = 8 a$, $N_x = 32$, $\alpha = 0.1$, and $\Gamma = 0.01$.  }
\label{Fig:Localtempphonon}
\end{figure}

We next attach to each lattice site and orbital a phonon bath with temperature $T_{\vec{r}_i}$. 
The temperature $T_{\vec{r}_i}$ varies in the real space; $T_{\vec{r}_i}$ is assumed to vary in the $\hat{x}$ direction, but to be constant in the $\hat{y}$ direction, $T_{\vec{r}_i} = T(x)$.
The coupling to the phonon baths can be described by the Hamiltonian 
\begin{align} \label{phononcoupling}
\hat{H}_{\rm{ph}}= \sum_{\vec{r}_i \sigma q} \left (g \Psi_{\vec{r}_i \sigma}^{\dagger} \Psi_{\vec{r}_i \sigma} \left (a^{\dagger}_{\vec{r}_i  \sigma  q} + a_{\vec{r}_i \sigma q } \right) + \omega_{q}
a^{\dagger}_{\vec{r}_i \sigma q}  a_{\vec{r}_i \sigma q}\right),
\end{align}
with electron-phonon coupling strength $g$.
Here $a^\dagger_{\vec{r}_i \sigma q}$ absorbs a phonon with energy $\omega_q$ and momentum $q$ attached to the point $\vec{r}_i$ and orbital $\sigma$ of the lattice. As considered before, we also attach to each lattice site a wire with 
a chemical potential $\mu_{\vec{r}_i}$ and the same temperature $T_{\vec{r}_i}$ as the phonon baths (see Eq.~\eqref{wirecoupling}). These wires are required to control the local chemical potentials of the Chern insulator.

The effect of the electron-phonon coupling is captured within Keldysh theory using a self-consistent one-loop approximation (i.e., the second order perturbation in $g$). Using that the system is translationally invariant in the $\hat y$ direction, we
introduce momenta $k_y$ but write all self-energies and Green functions as matrices in the $x$ coordinates (using either 16 or 32 sites with periodic boundary conditions). Below, we denote the diagonal elements of these matrices by $\Sigma(x)$ and $G(x)$.
The local lesser self-energy $\Sigma^{<}_{\rm{ph}}(x)$ and the imaginary part of the local retarded self-energy $\textrm{Im}\Sigma^R_{\rm{ph}} (x)$ are given by
\begin{widetext}
\begin{align} \label{appen:selfeq}
\Sigma^{<}_{\textrm{ph}} (x, \omega) &=- 2  g^2 \int \frac{d\omega_1}{(2 \pi)} b (\omega- \omega_1, x)\sum_{q}  \textrm{Im} D^R (x, q, \omega-\omega_1) \sum_{k_y} G^<(x, k_y, \omega_1), \nonumber \\ 
\textrm{Im} \Sigma^{R}_{\textrm{ph}} (x, \omega) &= - g^2 \int \frac{d\omega_1}{(2 \pi)} \sum_{q}
\textrm{Im} D^R (x, q, \omega-\omega_1) \sum_{k_y} \left ( \textrm{Im} G^< (x, k_y, \omega_1) - 2 b (\omega_1-\omega,x)\textrm{Im} G^R (x, k_y, \omega_1)\right). 
\end{align}
\end{widetext}
Here $b (\omega,x )$ is the Bose function at position $x$, i.e., $b (\omega,x ) = 1/ (\exp (\omega/ (k_B T(x))) - 1)$, and $D^{R}$ is the phonon retarded Green function. 
The real part of the retarded self energy is numerically obtained from the Kramers-Kronig relation, i.e., $\textrm{Re} \Sigma^{R}_{\textrm{ph}} (x, \omega) = P \int_{-\infty}^{\infty} d\omega' \textrm{Im} \Sigma^{R}_{\textrm{ph}} (x, \omega') / (\omega' - \omega) / \pi$. 
Furthermore, the phonon baths are assumed to follow the Ohmic behavior, i.e. $g^2 \sum_{q} \textrm{Im} D^R (x, q, \omega) \equiv \alpha \omega$ with dimensionless and spatially independent coupling constant $\alpha$. The Green functions can be written in terms of the self energies
\begin{align}\label{appen:greeneq}
G^{R} (k_y, \omega ) &= \left ( \left (g^{R} (k_y, \omega) \right )^{-1} - \Sigma^{R} (\omega) \right)^{-1},
\nonumber \\ 
G^{<} (k_y, \omega) &= G^{R} (k_y, \omega) \Sigma^{<} (\omega) G^{A} (k_y, \omega), 
\end{align}
with $\Sigma^R = \Sigma^{R}_{\textrm{ph}} + \Sigma^{R}_{\textrm{wire}}$, 
$\Sigma^< = \Sigma^{<}_{\textrm{ph}}  + \Sigma^{<}_{\textrm{wire}}$, and bare Green function
$g^{R} (k_y, \omega)$. 
The $\Sigma_{\textrm{wire}}$'s are the self-energies from the attached wires and can be written as $\Sigma_{\textrm{wire}}^{<} (\omega, x)= 2 i \Gamma f (\omega, x)$ and $\Sigma_{\textrm{wire}}^{R} (\omega)= - i \Gamma$ as shown in Eqs.~\eqref{appen:selfenergyretwire} and \eqref{appen:selfenergylesserwire}. Then, Green functions can be self-consistently calculated using Eqs.~\eqref{appen:selfeq} and \eqref{appen:greeneq}.

Our iteration scheme is as follows: For the initialization, the Green functions are obtained in the absence of the phonon baths. The Green functions are plugged into Eq.~\eqref{appen:selfeq} to obtain the self-energies, which are used to calculate the Green functions by matrix inversion employing Eq.~\eqref{appen:greeneq}. This procedure is repeated until the Green functions converge.
Convergence is typically reached by less than 10 iterations.


As specified in Sec.~\ref{sec:temperatureimprint}, provided that (i) all temperatures are much smaller than the gap $T(\vec r) \ll |M|$, the main contribution of the integral in Eq.~\eqref{currenttoThermometry} arises within the frequency range of $\omega \in (-T_0, T_0)$ inside the gap. In this frequency range, the retarded Green function decays on the length scale $\xi = v/|M|$ and hence the temperature at $\vec{r}$ is only sensitive to the temperatures in the vicinity of $\vec{r}$. Therefore, if (ii) $\xi \gg \sigma$, the temperature of the system is imprinted by the temperature profile of the phonons. At higher temperatures, when one has to consider the presence of thermally excited quasiparticles, one has also to make sure that $\sigma$ is large compared to the scattering length $\xi_{\rm ph}$ due to phonon scattering. For the plots shown in the paper, we focus, however, on the low-$T$ regime where thermal excitations can be ignored.

Those arguments above can be checked from the effective distribution function $f_{\vec{r}}^{\text{eff}} (\omega) $ of the Chern insulator, Eq.~\eqref{effectivedist}, obtained from computing the local Green's functions $G^< (x,x)$, $G^R (x, x)$, and $G^A (x, x)$, explicitly. $N_x = 32$ and $\sigma=8$ are used for this numerical simulation. In Fig.~\ref{Fig:Localtempphonon}, $f_{\vec{r}}^{\text{eff}} (\omega) $ is plotted as a function of $\omega$ with three different positions $x/\sigma = 0$ (green dots), $x/\sigma= 1$ (red dots), and $x/\sigma= 1.5$ (blue dots). $f_{\vec{r}}^{\text{eff}} (\omega) $ are in perfect agreement with the local Fermi-Dirac distribution function (solid lines) dictated by local phonon temperatures $T (x) = \overline{T}+T_0 \exp (-x^2/\sigma^2)$. Fig.~\ref{Fig:Localtempphonon}(b) shows the comparison of temperature profile $T(\vec{r})$ of the phonons (solid line) to the effective temperature $T^{\rm{eff}} (x)$ obtained from the attached thermometer. 
They are in good agreement, showing that the temperature of the system is imprinted by the temperature profile of the phonons. The slight deviation from the local phonon temperatures can be understood as a finite size effect. The smaller (larger) value at the center (edge) of the temperature profile implies that the temperature at position $\vec{r}$ is averaged over temperatures in the position range of $(\vec{r}-\sigma, \vec{r}+\sigma)$. A bigger system with larger $\sigma$ will suppress this deviation. 

\begin{widetext}
The charge current $J^{C}_{y} (x)$
and the energy current $J^{E}_{y} (x)$
along the $\hat{y}$ direction at position $x$ can be written from the same procedure specified in Sec.~\ref{app:Latticecalculation} as
\begin{align}
J^{C}_{y} (x) &= i e \int_{-\pi/a}^{\pi/a} \frac{d k_y}{2 \pi} \int \frac{d \omega}{2 \pi} 
 \textrm{Tr} 
\left [ \left (v \sigma_y \cos (k_y a) + \frac{2 \lambda^2}{a} \sigma_z \sin (k_y a) \right) G^{<} (x, k_y, \omega)  \right ]
\end{align}
and 
\begin{align}
J^{E}_{y} (x) =  - \frac{1}{\hbar a} \int_{-\pi/a}^{\pi/a}  \frac{d k_y}{2 \pi} \int \frac{d \omega}{2 \pi} 
&\bigg ( \left (2 \left (M + \frac{2 \lambda^2}{a^2} \right ) \lambda^2 \sin (k_y a) +  \left( \frac{\hbar^2 v^2}{2} - \frac{2 \lambda^4}{a^2} \right) \sin (2 k_y a) \right) \textrm{Tr} [ i G^{<} (x, k_y, \omega)] 
\nonumber \\ 
& + \frac{\hbar^2 v^2}{4}  \cos (k_y a) \textrm{Tr} \left [ \sigma_z  \textrm{Im} \left [ \left (G^{<} (x,x+a; k_y, \omega) - G^{<} (x,x-a; k_y, \omega)    \right) \right ] \right ] 
 \nonumber \\ & - \frac{\hbar \lambda^2 v}{2a} \sin (k_y a)  \textrm{Tr} \left [ \sigma_y  \textrm{Im} \left [ \left (G^{<}(x,x+a; k_y, \omega) - G^{<}(x,x-a; k_y, \omega)   \right) \right ] \right ]  \nonumber \\ 
 &- \frac{\hbar \lambda^2 v}{2a} \cos (k_y a)  \textrm{Tr} \left [ \sigma_x  \textrm{Re} \left [ \left (G^{<}(x,x+a; k_y, \omega)  + G^{<}(x,x-a; k_y, \omega)  \right) \right ] \right] \bigg ). 
\end{align}
\end{widetext}
Employing the lesser Green function obtained from the iteration scheme, we numerically obtain $J^{C}_{y} (x)$ and $J^{E}_{y} (x)$. $J^{C}_{y} (x)$ and $J^{E}_{y} (x)$ are plotted in Fig.~4 in the main text with the parameters written in the corresponding figure caption; $N_x = 16$ and $\sigma=4$ are used. The plot clearly shows that the Luttinger relation is invalid.  

We have thus obtained very similar results for a model where a temperature profile is imprinted by attached wire and a model where
it arises from the coupling to a phonon bath.
This  shows that the violation of the Luttinger relation is a generic feature of systems with spatially varying temperature profiles.
It is independent on how the temperature is induced and independent of whether the system is interacting or not.


\begin{thebibliography}{30}
\bibitem [{foo()}]{footnote}%
  \BibitemOpen
  \href@noop {} {}\bibinfo {note} {The heat current $\vec{J}^{H}$ can be
  generally written as $\vec{J}^{H} = \vec{J}^E - \mu \vec{J}^{C}$. We
  consider, however, $\mu=0$ only.}\BibitemShut {Stop}%
\end{thebibliography}
\end{document}